%% file: main.tex
\documentclass[10pt]{llncs}

\input{preamble}

\begin{document}

\maketitle

\input{macros}


\input{abs}

\input{intro}
\input{prelim}
\input{bmc}

\input{examples}

\input{qbf}
\input{cases}
\input{evals}
\input{related}
\input{concl}



\bibliographystyle{plain}
\bibliography{bibliography}

\appendix

\end{document}

%% file: preamble.tex
\mainmatter
\usepackage{tabulary}
\usepackage{fancyvrb}
\usepackage{multirow}
\usepackage{array}
\usepackage{booktabs}   
\usepackage{caption}
\usepackage{subcaption} 

\usepackage{amsmath,amssymb}
\usepackage[new]{old-arrows}
\usepackage{amsfonts}
\usepackage{xspace}
\usepackage[utf8]{inputenc}
\usepackage[T1]{fontenc}
\usepackage{array}
\usepackage{tabularx}
\usepackage{makecell}

\usepackage{paralist}
\usepackage{wrapfig}
\usepackage{listings}
\usepackage{tikz}

\usepackage{algorithmic}

\usepackage{todonotes}
\usepackage{ltl}

\tikzset{
    >=stealth,
    possible world/.style={circle,draw,thick,align=center},
    real world/.style={double,circle,draw,thick,align=center},
}
\usetikzlibrary{positioning,mindmap,automata,fit,backgrounds,shapes, 
arrows,shapes.misc,patterns}
\usetikzlibrary{decorations.pathmorphing}
\usetikzlibrary{decorations.markings}
\usetikzlibrary{calc}

\usepackage{graphicx}
\usepackage{tikz}
\usetikzlibrary{arrows, arrows.meta}
\usepackage[ruled,vlined,linesnumbered]{algorithm2e}
\usepackage{adjustbox}
\tikzset{arw/.style={>={Triangle[length=3mm,width=3mm]},line width=2mm,draw=gray}}

\title{Bounded Model Checking for Hyperproperties}
\author{Tzu-Han Hsu\inst{1} \and César Sánchez\inst{2} \and Borzoo 
Bonakdarpour\inst{1}}
\institute{
  Department of Computer Science and Engineering\\
  Michigan State University, USA\\
  \email{\{tzuhan,borzoo\}@msu.edu}
  \and
  IMDEA Software Institute, Spain  \\
  \email{cesar.sanchez@imdea.org}
}

\usepackage{cite}
\usepackage{stmaryrd} 

\usepackage{hyperref}
 \hypersetup{%
   colorlinks=true,           
   allcolors=blue!70!black,   
   pdfstartview=Fit,          
   breaklinks=true,
   pdfauthor={T.-H. Hsu, C. Sánchez, B. Bonakdarpour},
   pdftitle={Bounded Checking for Hyperproperties}}
  
    \tikzset{every picture/.style=semithick,
 }

%% file: macros.tex
\newcommand\modified[1]{#1}
\newcommand\bmodified[1]{#1}

\newcommand\frombaa[1]{\textcolor{red}{FROM BAA: #1}}
\newcommand\jyo[1]{\textcolor{blue}{Jyo: #1}}
\newcommand\chao[1]{\textcolor{blue}{Chao: #1}}
\newcommand\ufuk[1]{\textcolor{green}{Ufuk: #1}}
\newcommand\scott[1]{\textcolor{green}{Scott: #1}}
\newcommand\georgios[1]{\textcolor{brown}{Georgios: #1}}

\newcommand{\mypara}[1]{\vspace{0.5em} \noindent {\bf #1}.}
\newcommand{\myipara}[1]{\vspace{0.4em} \noindent {\em #1}.}

\newcommand{\lecps}{{\sc le}-{\sc cps}\xspace}
\newcommand{\lecpss}{{\sc le}-{\sc cps}\xspace}
\newcommand{\lecs}{{\sc lec}{\it s}\xspace}
\newcommand{\lec}{{\sc lec}\xspace}

\newcommand{\ignore}[1]{}
\newcommand{\halt}{\mathit{halt}}
\newcommand{\zplus}{\mathbb{Z}_{\geq 0}}

\newcommand\smin{\textcolor{red}{ins}}
\newcommand\smh{\textcolor{red}{hhs}}
\newcommand\smop{\textcolor{red}{ops}}
\newcommand\medin{\textcolor{red}{inm}}
\newcommand\medh{\textcolor{red}{hhm}}
\newcommand\medop{\textcolor{red}{opm}}
\newcommand\lgin{\textcolor{red}{inl}}
\newcommand\lgh{\textcolor{red}{hhl}}
\newcommand\lgop{\textcolor{red}{opl}}
\newcommand{\cur}{{\bf **}}
\newcommand{\current}{C}
\newcommand{\pending}{P}

\newcommand{\Phat}{\hat{P}}

\newcommand{\Paths}[2]{\mathit{Paths}^{#1}{(#2)}}
\newcommand{\fPaths}[2]{\mathit{Paths}^{#1}_{\mathit{fin}}{(#2)}}
\newcommand{\dom}{\mathit{dom}}
\newcommand{\dtmc}{\mathcal{M}}
\newcommand{\LTL}{\textsf{\small LTL}\xspace}
\newcommand{\CTL}{\textsf{\small CTL}\xspace}
\newcommand{\CTLstar}{\textsf{\small CTL$^*$}\xspace}
\newcommand{\PCTL}{\textsf{\small PCTL}\xspace}
\newcommand{\PCTLstar}{\textsf{\small PCTL$^*$}\xspace}
\newcommand{\HyperPCTL}{\textsf{\small HyperPCTL}\xspace}
\newcommand{\HyperLTL}{\textsf{\small HyperLTL}\xspace}
\newcommand{\HyperCTLstar}{\textsf{\small HyperCTL$^*$}\xspace}
\newcommand{\AFHyperLTL}{\mbox{AF-HyperLTL}\xspace}
\newcommand{\matching}{\mathcal{M}}
\newcommand{\topolgy}{\mathcal{T}}

\newcommand{\alphabet}{\mathrm{\Sigma}}
\newcommand{\states}{\mathrm{\Sigma}}
\newcommand{\statespace}{\states}
\newcommand{\Trace}{\mathsf{Traces}}
\newcommand{\trace}{t}
\newcommand{\qtrace}{\eta}
\newcommand{\sform}{\mathrm{\Phi}}
\newcommand{\pform}{\varphi}

\newcommand{\naturals}{\mathbb{N}_{>0}}
\newcommand{\naturalszero}{\mathbb{N}_{\geq 0}}

\newcommand{\AP}{\mathsf{AP}}
\newcommand{\Pos}{\mathsf{Pos}}

\newcommand{\Next}{\X}
\newcommand{\Finally}{\F}
\newcommand{\Globally}{\G}
\newcommand{\V}{\mathcal{V}}

\newcommand{\pr}{\mathbb{P}}
\renewcommand{\Pr}{\mathit{Pr}}

\newcommand{\emptyword}{\epsilon}

\newcommand{\init}{\mathit{init}}
\newcommand{\tpm}{\mathbf{P}}
\newcommand{\quant}{\mathbb{Q}}

\newcommand{\dbsim}{\mathit{dbSim}}
\newcommand{\res}{\mathit{res}}
\newcommand{\qout}{\mathit{qOut}}
\newcommand{\env}{\mathit{env}}
\newcommand{\fail}{\mathit{fail}}

\newcommand{\comp}[1]{\textsf{\small #1}}

\newcommand{\sigmakp}{$\mathsf{\Sigma^p_k}$\comp{-complete}\xspace}
\newcommand{\pikp}{$\mathsf{\Pi^p_k}$\comp{-complete}\xspace}

\newcommand\donotshow[1]{}

\newcommand{\ie}{i.e.\xspace}
\newcommand{\shield}{SHIELD\xspace}
\newcommand{\lestl}{\leccomp[{\sc stl}]}
\newcommand{\mulf}[1]{\multicolumn{2}{l}{#1}}

\newcommand{\z}{\cellcolor{black}}
\newcommand{\x}{\cellcolor{lightgray}}

\newcommand{\code}[1]{\textsf{\small #1}}

\newcommand{\tru}{\mathsf{true}}
\newcommand{\fals}{\mathsf{false}}
\newcommand{\inn}{\mathsf{in}}
\newcommand{\out}{\mathsf{out}}
\newcommand{\suffix}[3]{#1[#2,#3]}
\newcommand{\F}{\LTLdiamond}
\newcommand{\always}{\LTLsquare}
\newcommand{\G}{\LTLsquare}
\newcommand{\until}{\mathbin{\mathcal{U}}}
\newcommand{\U}{\until}
\newcommand{\release}{\mathbin{\mathcal{R}}}
\newcommand{\W}{\,\mathcal W\,}
\newcommand{\X}{\LTLcircle}

\newcommand{\qbf}[1]{\llbracket #1 \rrbracket}
\newcommand{\States}{S}
\newcommand{\state}{s}
\newcommand{\trans}{\delta}
\newcommand{\inp}{\mathfrak{u}}
\newcommand{\uncont}{\mathfrak{u}}
\newcommand{\cont}{\mathfrak{c}}
\newcommand{\outp}{\mathcal{C}}
\newcommand{\kframe}{\mathcal{F}}
\newcommand{\krip}{K}
\newcommand{\Kr}{\mathcal{K}}
\newcommand{\Tr}{\mathcal{T}}
\newcommand{\ktuple}{\langle S, S_\init, \trans, L \rangle}
\newcommand{\ktupleprime}{\langle S', S'_\init, \cont', \uncont', L' \rangle}
\newcommand{\ktupleprimewithuc}{\langle S', S'_\init, \trans' \cup \inp, L' 
\rangle}
\newcommand{\lang}{\mathcal{L}}

\newcommand{\pos}{\mathit{pos}}
\newcommand{\negt}{\mathit{neg}}
\newcommand{\CS}[2]{\mbox{\sf \small CS[#1, #2{}]}\xspace}

\newcolumntype{K}[1]{>{\centering\arraybackslash}p{#1}}

\newcommand{\dbr}[1]{\llbracket #1 \rrbracket}

\newcommand{\GMNI}{\textsf{\small GMNI}\xspace}
\newcommand{\GNI}{\textsf{\small GNI}\xspace}

\pgfdeclarelayer{background}
\pgfdeclarelayer{foreground}
\pgfsetlayers{background,main,foreground}

\tikzset{
  invisible/.style={opacity=0, text opacity=0},
  visible on/.style={alt={#1{}{invisible}}},
  alt/.code args={<#1>#2#3}{%
    \alt<#1>{\pgfkeysalso{#2}}{\pgfkeysalso{#3}} 
the path
  },
}

\newcommand\enrec[1]{%
  \tikz[baseline=(X.base)]
    \node (X) [draw, shape=circle, inner sep=0, fill=white] {\strut #1};}

\newcommand\encirclew[1]{%
  \tikz[baseline=(X.base)]
    \node (X) [draw, shape=circle, inner sep=0, fill=white] {\strut #1};}

\newcommand\encirclegr[1]{%
  \tikz[baseline=(X.base)]
    \node (X) [draw, shape=circle, inner sep=0, fill=gray] {\strut #1};}

 \newcommand\encirclep[1]{%
  \tikz[baseline=(X.base)]
    \node (X) [draw, shape=circle, inner sep=0, fill=pink] {\strut #1};}

\newcommand\encircle[1]{%
  \tikz[baseline=(X.base)]
    \node (X) [draw, shape=circle, inner sep=0, fill=green] {\strut #1};}

\newcommand\encircley[1]{%
  \tikz[baseline=(X.base)]
    \node (X) [draw, shape=circle, inner sep=0, fill=yellow] {\strut #1};}

\newcommand\encircler[1]{%
  \tikz[baseline=(X.base)]
    \node (X) [draw, shape=circle, inner sep=0, fill=red] {\strut #1};}

\tikzstyle{place}=[circle,thick,draw=blue!75,fill=blue!20,minimum size=6mm]
\tikzstyle{red place}=[place,draw=red!75,fill=red!20]
\tikzstyle{transition}=[rectangle,thick,draw=black, fill=black, 
minimum size=1mm]

\newcommand{\dec}{\mathtt{dec}}
\newcommand{\ses}{\mathtt{ses}}
\newcommand{\session}{\mathtt{session}}
\newcommand{\status}{\mathtt{status}}
\newcommand{\ntf}{\mathtt{ntf}}

\definecolor{mGreen}{rgb}{0,0.6,0}
\definecolor{mGray}{rgb}{0.5,0.5,0.5}
\definecolor{mPurple}{rgb}{0.58,0,0.82}
\definecolor{backgroundColour}{rgb}{0.95,0.95,0.92}

\lstdefinestyle{CStyle}{
    backgroundcolor=\color{backgroundColour},   
    commentstyle=\color{mGreen},
    keywordstyle=\color{magenta},
    numberstyle=\tiny\color{mGray},
    stringstyle=\color{mPurple},
    basicstyle=\footnotesize,
    breakatwhitespace=false,         
    breaklines=true,                 
    captionpos=b,                    
    keepspaces=true,                 
    numbers=left,                    
    numbersep=2pt,                  
    showspaces=false,                
    showstringspaces=false,
    showtabs=false,                  
    tabsize=2,
    language=C
}


\newcommand{\eab}[1]{{\color{red}#1}}

\renewcommand{\topfraction}{0.96}
\renewcommand{\bottomfraction}{0.95}
\renewcommand{\textfraction}{0.1}
\renewcommand{\floatpagefraction}{1}
\renewcommand{\dbltopfraction}{.97}
\renewcommand{\dblfloatpagefraction}{.99}

\newcommand{\tupleof}[1]{\langle#1\rangle}
\newcommand{\OR}{\vee}
\newcommand{\AND}{\wedge}

\newcommand{\cesar}[1]{\todo[linecolor=red,backgroundcolor=red!25,bordercolor=red]{Cesar: #1}}
\newcommand{\tzuhan}[1]{\todo[linecolor=pink,backgroundcolor=pink!20,bordercolor=pink]{Tzu-Han: #1}}
\newcommand{\borzoo}[1]{\todo[linecolor=green,backgroundcolor=green!25,bordercolor=green]{Borzoo: #1}}

\newcommand{\PR}[1]{\ensuremath{\textit{#1}}}
\newcommand{\Halt}{\PR{halt}\xspace}
\newcommand{\Halted}{\PR{halted}\xspace}
\newcommand{\AllHalt}{\ensuremath{\bigwedge_{\pi \Vars(\varphi)}\Halt_\pi}}
\newcommand{\DefinedAs}{\,\stackrel{\text{def}}{=}\,}

\newcommand{\OPT}{\mathit{opt}}
\newcommand{\PES}{\mathit{pes}}
\newcommand{\HOPT}{\mathit{hopt}}
\newcommand{\HPES}{\mathit{hpes}}

\newcommand{\Pmodels}{\models^{\PES}}
\newcommand{\Omodels}{\models^{\OPT}}
\newcommand{\HOmodels}{\models^{\HOPT}}
\newcommand{\HPmodels}{\models^{\HPES}}
\newcommand{\Hmodels}{\models^{\textit{halt}}}

\newcommand{\Pkmodels}{\models_k^{\PES}}
\newcommand{\Okmodels}{\models_k^{\OPT}}
\newcommand{\HOkmodels}{\models_k^{\HOPT}}
\newcommand{\HPkmodels}{\models_k^{\HPES}}

\newcommand{\HP}{\textit{HP}}
\newcommand{\HO}{\textit{HO}}

\renewcommand{\And}{\mathrel{\wedge}}
\newcommand{\Or}{\mathrel{\vee}}

\newcommand{\HyperQube}{\code{HyperQube}\xspace}
\newcommand{\Ocaml}{\code{Ocaml}\xspace}
\newcommand{\select}{\mathit{select}}
\newcommand{\sym}{\mathit{sym}}
\newcommand{\pause}{\mathit{pause}}
\newcommand{\pc}{\mathit{pc}}
\newcommand{\pushRight}{\mathit{pushRight}}
\newcommand{\popLeft}{\mathit{popLeft}}
\newcommand{\history}{\mathit{history}}
\newcommand{\invocations}{\mathit{invocations}}
\newcommand{\response}{\mathit{responses}}
\newcommand{\linearizability}{\mathit{lin}}

\newcommand{\rh}{\textit{\rh}}
\newcommand{\RH}{\textit{\RH}}
\newcommand{\lh}{\textit{\lh}}
\newcommand{\LH}{\textit{\LH}}

\newcommand{\NI}{\mathit{NI}}
\newcommand{\terminate}{\mathit{terminate}}
\newcommand{\PIN}{\mathit{PIN}}
\newcommand{\Result}{\mathit{Result}}

\newcommand{\NRR}{\mathit{NRR}}
\newcommand{\NRO}{\mathit{NRO}}

\newcommand{\partyA}{P}
\newcommand{\partyB}{Q}
\newcommand{\Act}{\mathit{Act}}
\newcommand{\skipp}{\mathit{skip}}

\newcommand{\effectiveness}{\mathit{effectiveness}}
\newcommand{\fairnessforA}{\mathit{fairness\ for\ A}}
\newcommand{\fairnessforB}{\mathit{fairness\ for\ B}}

\newcommand{\goal}{\mathit{goal}}

\newcommand{\strategy}{\mathit{strategy}}

\newcommand{\correct}{\mathit{correct}}
\newcommand{\incorrect}{\mathit{incorrect}}

\newcommand{\fair}{\mathit{fair}}

\newcommand{\mut}{\mathit{mut}}
\newcommand{\inputt}{\mathit{in}}
\newcommand{\outputt}{\mathit{out}}

\newcommand{\hpes}{\ensuremath{\mathit{halting-pessimistic}}}
\newcommand{\hopt}{\ensuremath{\mathit{halting-optimistic}}}

\newcommand{\pes}{\textit{pessimistic}}
\newcommand{\opt}{\textit{optimsitic}}
\newcommand{\bpes}{\textit{halting-pes}}
\newcommand{\bopt}{\textit{halting-opt}}
\newcommand{\shortest}{\mathit{shortest\ path}}
\newcommand{\rub}{\mathit{robustness}}

\newcommand{\QBFall}[4]{\qbf{#4}^{#1}_{#2,#3}}
\newcommand{\QBFany}[3]{\QBFall{*}{#1}{#2}{#3}}
\newcommand{\QBF}[3]{\QBFall{}{#1}{#2}{#3}}
\newcommand{\QBFpes}[3]{\QBFall{\PES}{#1}{#2}{#3}}
\newcommand{\QBFopt}[3]{\QBFall{\OPT}{#1}{#2}{#3}}
\newcommand{\QBFhpes}[3]{\QBFall{\HPES}{#1}{#2}{#3}}
\newcommand{\QBFhopt}[3]{\QBFall{\HOPT}{#1}{#2}{#3}}

\newcommand{\Vars}{\textit{Vars}}
\newcommand{\Traces}{\textit{Traces}}

\newcommand{\AT}{\textit{at}}

\newcommand{\so}{\textit{ thus }}

%% file: abs.tex
%
%
%
%
%

\begin{abstract}
  Hyperproperties are properties of systems that relate multiple
  computation traces, including security properties and properties in
  concurrency.
  This paper introduces a bounded model checking (BMC) algorithm for
  hyperproperties expressed in HyperLTL, which---to the best of our
  knowledge---is the first such algorithm.
  Just as the classic BMC technique for LTL primarily aims at finding
  bugs, our approach also targets identifying counterexamples.
  LTL describes a property via inspecting individual traces, so BMC for
  LTL is reduced to SAT solving.
  HyperLTL allows explicit and simultaneous quantification over traces
  and describes properties that involves multiple traces and, hence,
  our BMC approach naturally reduces to QBF solving.
  We report on successful and efficient model checking, implemented in
  a tool called \HyperQube, of a rich set of experiments on a variety
  of case studies, including security/privacy, concurrent data
  structures, and path planning in robotics applications.
\end{abstract}


%% file: intro.tex
\section{Introduction}
\label{sec:intro}

{\em Hyperproperties}~\cite{cs10} have been shown to be a powerful framework
for specifying and reasoning about important classes of requirements
that were not possible with trace-based languages such as the classic
temporal logics.
Examples include information-flow security, consistency
models in concurrent computing~\cite{bss18}, and robustness models in
cyber-physical systems~\cite{wzbp19,bonakdarpour20model}.
The temporal logic HyperLTL~\cite{cfkmrs14} extends LTL by allowing
explicit and simultaneous quantification over execution traces,
describing the property of multiple traces.
For example, the security policy {\em observational determinism} can
be specified by the following HyperLTL formula:
\[
\forall \pi_A.\forall \pi_B.(o_{\pi_A} \leftrightarrow o{_{\pi_B}}) \, \W \, \neg (i_{\pi_A} \leftrightarrow i{_{\pi_B}}) 
\]
which stipulates that every pair of traces $\pi_A$ and $\pi_B$ have to agree on
the value of the (public) output $o$ as long as they agree on the
value of the (secret) input $i$, where `$\W$' denotes the weak until
operator.

There has been a recent surge of model checking techniques for
HyperLTL specifications~\cite{cfst19,fht18,frs15,cfkmrs14}.
These approaches employ various techniques (e.g., alternating
automata, model counting, strategy synthesis, etc) to verify
hyperproperties.
However, they generally fall short in proposing an effective method to
deal with identifying bugs with respect to alternating HyperLTL
formulas.
Indeed, quantifier alternation has been shown to generally elevate the
complexity class of model checking HyperLTL specifications in
different shapes of Kripke structures (KS)~\cite{bf18,cfkmrs14}.
For example, consider the simple Kripke structure $\krip$ in
Fig.~\ref{fig:kripke} and HyperLTL formulas
$\varphi_1 = \forall \pi_A.\forall \pi_B. \G(p_{\pi_A} \leftrightarrow
p_{\pi_B})$ and
$\varphi_2 = \forall \pi_A.\exists \pi_B. \G(p_{\pi_A} \not \leftrightarrow
p_{\pi_B})$.
Proving that $\krip \not \models \varphi_1$ (where traces for $\pi_A$
and $\pi_B$ are taken from $K$) can be reduced to building the
self-composition of $\krip$ and applying standard LTL model checking,
resulting in worst-case complexity $|\krip|^2$ in the size of the
system.
On the contrary, proving that $\krip \models \varphi_2$ is not as
straightforward.
In the worst case, this requires a subset generation to encode the
existential quantifier within the Kripke structure, resulting in
$|\krip|\cdot 2^{|\krip|}$ blow up.
In addition, the quantification is over traces rather than states,
adding to the complexity of reasoning.

\input{figs/ks}

Following the great success of bounded model checking (BMC) for LTL
specifications~\cite{cbrz01}, in this paper, we propose the first
BMC algorithm for HyperLTL.
To the best of our knowledge this is the first such algorithm.
Just as BMC for LTL is reduced to SAT solving to search for a
counterexample trace whose length is bounded by some integer $k$, we
reduce BMC for HyperLTL to QBF solving to be able to deal with
quantified counterexample traces in the input model.
More formally, given a HyperLTL formula (for example, of the form)
$\varphi = \forall \pi_A. \exists \pi_B.\psi$ and a family of Kripke
structures $\Kr = (\krip_A, \krip_B)$ (one per trace
variable), the reduction involves three main components.
First, the transition relation of $\krip_\pi$ (for every $\pi$) is
represented by a Boolean encoding $\qbf{\krip_\pi}$.
Secondly, the inner LTL subformula $\psi$ is translated to a Boolean
fixpoint representation $\qbf{\psi}$ in a similar fashion to the
standard BMC technique for LTL.
This way, the QBF encoding for a bound $k \geq 0$ roughly appears as:
\begin{equation}
  \label{eq:qbfenc}
\qbf{\Kr, \neg \varphi}_k = \exists \overline{x_A}.\forall \overline{x_B}.\qbf{\krip_A}_k
\wedge \big(\qbf{\krip_B}_k \rightarrow \qbf{\neg \psi}_k\big)
\end{equation}
where the vector of Boolean variables $\overline{x_A}$ (respectively,
$\overline{x_B}$) are used to represent the state and propositions of
the kripke structures $K_A$ (resp. $K_B$) for steps from $0$ to $k$.
Formulas $\qbf{\krip_A}_k$ and $\qbf{\krip_B}_k$ are the unrollings
$K_A$ (using $\overline{x_A}$) and $K_B$ (using $\overline{x_B}$), and
  $\qbf{\neg\psi}$ (that uses both $\overline{x_A}$ and $\overline{x_B}$)
  is the fixpoint Boolean encoding of $\neg\psi$.
We note that the proposed technique in this paper does not incorporate
a loop condition, as implementing such a condition for multiple traces
is not straightforward at all.
This, of course, comes at the cost of lack of a completeness result.
%

While our QBF encoding is a natural generalization of BMC for
HyperLTL, the first contribution of this paper is a more refined view
of how to interpret the behavior of the formula beyond the unrolling
depth $k$.
Consider LTL formula $\forall \pi.\G p_\pi$.
BMC for LTL attempts to find a counterexample by unrolling the model
and check for satifiability of $\exists \pi.\F \neg p_\pi$.
In this case satifiability means existence of a counterexample within
the first $k$ steps.
Now consider LTL formula $\forall \pi.\F p_\pi$ whose negation is of
the form $\exists \pi.\G \neg p_\pi$.
In the classic BMC, due to its {\em pessimistic} handling of $\G$ the
unstatisfiability of the formula can not be established in the finite
unrolling (handling these formulas requires to appeal to either
looping conditions or to reach the diameter of the the system).
This is because $\G\neg p_\pi$ is not \emph{sometimes finitely
  satisfiable} (SFS), in the terminology introduced by Havelund and
Peled~\cite{hp18}, meaning that not all satisfying traces of
$\G p_\pi$ have a finite prefix that witness the satisfiability.

We propose a method that allows to interpret a wide range of outcomes
of the QBF solver and relate these to the original model checking
decision problem.
To this end, we propose the following semantics for BMC for HyperLTL:
\begin{itemize}
\item \emph{Pessimistic} semantics (which is the common for LTL BMC)
  under which pending eventualities are considered to be unfulfilled.
  This semantics work for sometime finitely satisfiable temporal
  formulas and paves the way for bug hunting.
\item \emph{Optimistic} semantics considers the dual case, where
  pending eventualities are assumed to be fulfilled at the end of the
  trace.
  This semantics work for \emph{sometimes finitely refutable} formulas,
  and allows us to interpret unsatisfiability of QBF as proof of
  verification even with bounded traces.
\item \emph{Halting} variants of the optimistic and pessimistic
  semantics, which allows sound and complete decision on a verdict for
  terminating models.
\end{itemize}  

We have fully implemented our technique in the tool \HyperQube.
Our experimental evaluation includes a rich set of case studies, such
as information-flow security/privacy, concurrent data structures (in
particular, linearizability), and path planning in robotic
applications.
Our evaluation shows that our technique is effective and efficient in
identifying bugs in several prominent examples. We also show that our
QBF-based approach is certainly more efficient than an brute-force
SAT-based approach, where universal and existential quantifiers are
eliminated by combinatorial expansion to conjunctions and disjunctions.
We also show that in some cases our approach can also be used as as tool for
synthesis.
Indeed, a witness to an existential quantifier in a HyperLTL formula
is an execution path that satisfies the formula.
For example, our experiments on path planning for robots showcases this
feature of \HyperQube.

In summary, the contributions of this paper are as follows:

\begin{itemize}
\item We propose a QBF-based BMC approach for verification and
  falsification of HyperLTL specifications.
\item We introduce complementary semantics that allow proving and
  disproving formulas, given a finite set of finite traces.
\item We rigorously analyze the performance of our technique by case
  studies from different areas of computing.
\end{itemize}

The rest of the paper is structured as
follows.
Section~\ref{sec:prelim} contains the preliminaries.
Section~\ref{sec:bmc} introduces the different bounded semantics for
HyperLTL.
Section~\ref{sec:qbf} presents the encoding into QBF of the different
formulas and bounded semantics of choice, and what can be inferred in
each case about the HyperLTL model checking problem in each case.
Sections~\ref{sec:cases} and~\ref{sec:evals} present an empirical
evaluation of our tool \code{HyperQube}. Section~\ref{sec:related} presents
the related work and Section~\ref{sec:concl} concludes.


%% file: figs/ks.tex
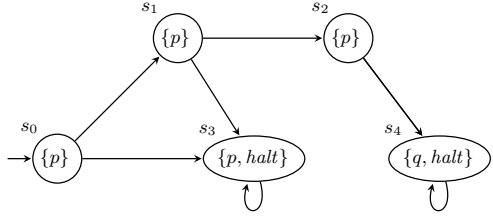
\begin{figure}[]
\centering
\scalebox{0.8}{
\begin{tikzpicture}[-,>=stealth,shorten >=.5pt,auto,node distance=2cm, 
semithick, initial text={}]

\node[initial, state] [text width=1em, text centered, minimum 
 height=2.0em](0) at (0,0) {\hspace*{-1.25mm}$\{p\}$};
\node [above left = 0.005 cm and -0.1 cm of 0](label){$s_0$};
\node[state, above right=of 0][text width=1em, text centered, minimum 
 height=2.5em] (1) {\hspace*{-1.25mm}$\{p\}$};
\node [above left = 0.005 cm and -0.1 cm of 1](label){$s_{1}$};
\node[state, right=of 1][text width=1em, text centered, minimum 
height=2.5em] (2){\hspace*{-1.25mm}$\{p\}$};
\node [above left = 0.005 cm and -0.1 cm of 2](label){$s_{2}$};

\node[state, right=of 0][text width=3em, text centered, ellipse, minimum height=1em, minimum width = 1em] (3) {\hspace*{-1.25mm}$\{p,\halt\}$};
\node [above left = 0.005 cm and -0.1 cm of 3](label){$s_{3}$};

\node[state, right=of 3][text width=3em, text centered, ellipse, minimum height=1em, minimum width = 1em] (4) at (3.5,0) {\hspace*{-1.25mm}$\{q,\halt\}$};
\node [above left = 0.005 cm and -0.1 cm of 4](label){$s_{4}$};

\draw[->]  
 (0) edge node (01 label) {} (1)
 (1) edge node (12 label) {} (3)
 (0) edge node (03 label) {} (3)
 (1) edge node (12 label) {} (2)
 (2) edge node (24 label) {} (4)
 (2) edge node (24 label) {} (4)
 (3) edge [loop below] node (22 label) {} (3)
 (4) edge [loop below] node (33 label) {} (4);
 
 \end{tikzpicture}
}
\caption {A Kripke structure.}
\label{fig:kripke}
\end{figure}

%% file: prelim.tex
\section{Preliminaries}
\label{sec:prelim}


\subsection{Kripke Structures}
\label{subsec:krip}

Let $\AP$ be a finite set of {\em atomic propositions} and $\alphabet = 
2^{\AP}$ be the {\em alphabet}. A {\em letter} is an element of $\alphabet$. A 
\emph{trace} $t\in \Sigma^\omega$ over alphabet $\alphabet$ is an infinite sequence of letters: $t = t(0)t(1)t(2) \cdots$


\begin{definition}
\label{def:kripke}
A {\em Kripke structure} is a tuple $\krip = \ktuple$,
where 

\begin{itemize}
 \item $\States$ is a finite set of {\em states};
 \item $\States_{\init} \subseteq \States$ is the set of {\em initial states};
 \item $\trans \subseteq \States \times \States$ is a {\em transition 
relation}, and 
 \item $L: S \rightarrow \statespace$ is a {\em labeling function} on the 
states of $\krip$.
\end{itemize}
We require that 
for each $\state \in \States$, there exists $\state' \in \States$, such 
that $(\state, \state') \in \trans$.

\end{definition}

Fig.~\ref{fig:kripke} shows a Kripke structure, where
$\States_{\init} = \{s_0\}$, $L(s_0)= \{p\}$, $L(s_4)=\{q, halt\}$, etc.

The \emph{size} of the Kripke structure is the number of its states.

%
A {\em loop} in 
$\krip$ is a finite sequence $\state(0)\state(1)\cdots \state(n)$, such that 
$(\state(i), \state({i+1})) \in \trans$, for all $0 \leq i < n$, and $(\state(n), 
\state(0)) \in \trans$. We call a Kripke frame {\em acyclic}, if the only loops 
are self-loops on otherwise terminal states, i.e., on states that have no other outgoing 
transition. Since Definition~\ref{def:kripke} does not allow terminal states, we only
consider acyclic Kripke structures with such added self-loops. We also label
such states by atomic proposition $\halt$.


A \emph{path} of a Kripke structure is an infinite sequence of states
$\state(0)\state(1)\cdots \in \States^\omega$, such that:

\begin{itemize}
 \item $\state(0) \in \States_\init$, and
\item $(\state(i), \state({i+1})) \in \trans$, for all $i \geq 0$. 
\end{itemize}
A trace of a Kripke structure is a trace
$t(0)t(1)t(2) \cdots \in \alphabet^\omega$, such that there exists a
path $\state(0)\state(1)\cdots \in \States^\omega$ with
$t(i) = L(\state(i))$ for all $i\geq 0$. We denote by
$\Traces(\krip, \state)$ the set of all traces of $\krip$ with paths
that start in state $\state \in \States$, and use $\Traces(\krip)$ as
a short for $\bigcup_{s \in \States_{\init}}\Traces(\krip,\state)$.


\subsection{The Temporal Logic HyperLTL}
\label{subsec:hltl}

HyperLTL~\cite{cfkmrs14} is an extension of the linear-time temporal logic
(LTL) for hyperproperties.
The syntax of HyperLTL formulas is defined inductively by the following grammar:
\begin{equation*}
\begin{aligned}
& \varphi ::= \exists \pi . \varphi \mid \forall \pi. \varphi \mid \phi \\
& \phi ::= \tru \mid a_\pi \mid \lnot \phi \mid \phi \OR \phi \mid \phi \AND \phi
             \mid \phi \until \, \phi \mid \phi \release \, \phi \mid \X \phi
    \end{aligned}
\end{equation*}
where $a \in \AP$ is an atomic proposition and $\pi$ is a {\em trace variable}
from  an infinite supply of variables $\V$.
The Boolean connectives $\neg$, $\OR$ and $\AND$ have the usual
meaning, $\until$ is the temporal \emph{until} operator, $\release$ is
the temporal \emph{release} operator, and $\X$ is the temporal
\emph{next} operator.
We also consider other derived Boolean connectives, such as
$\rightarrow$, and $\leftrightarrow$, and the derived temporal
operators \emph{eventually} $\F\varphi\equiv \tru\,\until\varphi$ and
\emph{globally} $\always\varphi\equiv\neg\F\neg\varphi$.
Even though the set of operators presented is not minimal, we have
introduced this set to uniform the treatment with the variants in
Section~\ref{sec:boundedHyperLTL}.
The quantified formulas $\exists \pi$ and $\forall \pi$ are read as
``along some trace $\pi$'' and ``along all traces $\pi$'',
respectively.
A formula is {\em closed} (i.e., a {\em sentence})  if all trace
variables used in the formula are quantified.
We assumed, without lost of generality that no variable is quantified
twice.
We use $\Vars(\varphi)$ for the set of path variables used in
formula $\varphi$.

\paragraph{Semantics.}

An interpretation $\Tr=\tupleof{T_\pi}_{\pi\in\Vars(\varphi)}$ of a
formula $\varphi$ consists of a set of traces, one set $T_\pi$ per
trace variable $\pi$ in $\Vars(\varphi)$.
We use $T_\pi$ for the set of traces assigned to $\pi$.
The idea here is to allow quantifiers to range over different models.
We will use this feature in the verification of hyperproperties such
as linearizabiliity, where different quantifiers are associated with
different sets of executions (in this case one for the concurrent
implementation and one for the sequential implementation).
That is, each set of traces comes from a Kripke structure and we use
$\Kr=\tupleof{K_\pi}_{\pi\in\Vars(\varphi)}$ to denote a {\em family} of
Kripke structure, so $T_\pi=\Traces(K_\pi)$ is the traces that $\pi$
can range over, which comes from $K_\pi$.
Abusing notation, we write $\Tr=\Traces(\Kr)$.

Note that all trace sets being the same set of traces for a single
Kripke structure $K$ (i.e. $K_\pi=K$ for all $\pi$) is a particular
case, which leads to the original semantics of HyperLTL~\cite{cfkmrs14}.
The semantics of HyperLTL are defined with respect to a trace
assignment, which is a partial
map~$\Pi \colon \Vars(\varphi) \rightharpoonup\alphabet^\omega$.
The assignment with empty domain is denoted by $\Pi_\emptyset$.
Given a trace assignment~$\Pi$, a trace variable~$\pi$, and a concrete
trace~$t \in \alphabet^\omega$, we denote by $\Pi[\pi \rightarrow t]$
the assignment that coincides with $\Pi$ everywhere but at $\pi$,
which is mapped to trace $t$.

The satisfaction of a HyperLTL formula $\varphi$ is a binary relation
$\models$ that associates a formula to the models $(\Tr,\Pi,i)$ where
$i \in \zplus$ is a pointer that indicates the current position of all
traces in $\Tr$.
The semantics is defined as follows:
\[
  \begin{array}{l@{\hspace{1em}}c@{\hspace{1em}}l}
  (\Tr, \Pi,0) \models \exists \pi.\ \modified{\psi} & \text{iff} & \text{ there is a  } t \in T_\pi \text{ such that } (\Tr,\Pi[\pi \rightarrow t],0) \models \psi,\\
    (\Tr, \Pi,0) \models \forall \pi.\ \modified{\psi} & \text{iff} & \text{ for all } t \in T_\pi \text{ such that } (\Tr,\Pi[\pi\rightarrow t],0) \models \psi,\\
  (\Tr, \Pi,i) \models \tru \\
  (\Tr, \Pi,i) \models a_\pi & \text{iff} & a \in \Pi(\pi)(i),\\
  (\Tr, \Pi,i) \models \neg \psi & \text{iff} & (\Tr, \Pi,i) \not\models \psi,\\
  (\Tr, \Pi,i) \models \psi_1 \OR \psi_2 & \text{iff} & (\Tr, \Pi,i) \models \psi_1\text{ or } (\Tr, \Pi,i) \models \psi_2,\\
  (\Tr, \Pi,i) \models \psi_1 \AND \psi_2 & \text{iff} & (\Tr, \Pi,i) \models \psi_1
\text{ and } (\Tr, \Pi,i) \models \psi_2,\\
  (\Tr, \Pi,i) \models \X \psi & \mbox{iff} & (\Tr,\Pi,i+1)\models\psi,\\
  (\Tr, \Pi,i) \models \psi_1 \until \psi_2 & \text{iff} &  \text{there is a } j \ge i \text{ for which } (\Tr,\Pi,j) \models \psi_2 \text{ and } \\
 && \hspace{1em} \text{for all } k \in [i, j), (\Tr,\Pi,k)\models \psi_1,\\
  (\Tr, \Pi,i) \models \psi_1 \release \psi_2 & \text{iff} &  \text{either for all } j \geq i,\;  (\Tr,\Pi,j) \models \psi_2 \text{, or, } \\
&& \hspace{1em} \text{for some } j\geq i, (\Tr,\Pi,j)\models \psi_1 \text{ and }\\
&& \hspace{1em} \text{for all }  k\in [i, j]: (\Tr,\Pi,k)\models\psi_2.
  \end{array}
\]

We say that an interpretation $\Tr$ satisfies a sentence~$\varphi$,
denoted by $\Tr \models \varphi$, if
$(\Tr, \Pi_\emptyset,0) \models \varphi$.
We say that a family of Kripke structures $\Kr$ satisfies a sentence~$\varphi$,
denoted by $\Kr \models \varphi$, if $\langle \Traces(\krip_\pi)\rangle_{\pi \in \Vars(\varphi)} \models \varphi$.
When the same kripke structure $K$ is used for all path variables we
write $K\models\varphi$.

For example, the Kripke structure in Fig.~\ref{fig:kripke} satisfies
HyperLTL formula
$\varphi = \forall\pi_A.\exists \pi_B . \F(p_{\pi_A} \leftrightarrow
q_{\pi_B})$.

These semantics are slightly different from the definition
in~\cite{cfkmrs14}, but equivalent.
First, we use the pointer $i$ instead of chopping the trace with the
head elements when traversing the tuple of traces forward, but this is
clearly equivalent and more convenient later in the paper when we
define finite unrollings.
Second, we use a multi-model semantics allowing different trace variables to
choose traces from different trace sets.
In terms of the model checking problem, multi-model and (the
conventional) single-model semantics~\cite{cfkmrs14} are equivalent.
One can instantiate all models with the same Kripke structure (so
multi-model can simulate single-model).
For the other direction, one can merge all Kripke structures into a
single Kripke structure and add fresh predicates to distinguish each
Kripke structure, and then require each path to belong to the desired
original Kripke structure.

\subsection{Quantified Boolean Formula Satisfiability}

\begin{figure}
\centering
\includegraphics[scale=.9]{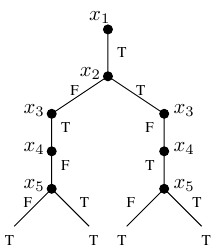}
\caption{Model for the QBF formula.}
\label{fig:qbfmodel-sys}
\end{figure}

The {\em quantified Boolean formula} (QBF) satisfiability problem~\cite{gj79} 
is the following:

\begin{quote}
 
{\em Given is a set of Boolean variables, $\{x_1, x_2, \dots, x_n\}$, and a 
quantified Boolean formula $F=\quant_1 x_1.\quant_2 x_2\dots\quant_{n-1} x_{n-1}.\quant_n x_n.\psi$,
where each $\quant_i \in \{\forall, \exists\}$ \textup{(}$i \in [1, n]$\textup{)} and $\psi$ is
an arbitrary Boolean formula over variables $\{x_1, \ldots, x_n\}$. Is $F$ 
true?}

\end{quote}
Solving the satisfiability problem for QBF is known to be
PSPACE-complete.  Figure~\ref{fig:qbfmodel-sys} shows a satisfying
model for the following formula:
\begin{align*}
F = \exists x_1.\forall x_2.\exists x_3.\exists x_4.\forall x_5. & (x_1 \vee \neg 
x_2 \vee x_3) \; \wedge \; (\neg x_1 \vee x_2 \vee \neg x_4) \; \wedge \\
& (\neg x_3 \vee x_4 \vee \neg x_5) \; \wedge \; (x_1 \vee x_4 \vee x_5).
\end{align*}

%% file: bmc.tex
\section{Bounded Semantics for HyperLTL}
\label{sec:boundedHyperLTL}
\label{sec:bmc}

\newcolumntype{C}{>{$}c<{$}}
\newcolumntype{L}{>{$}l<{$}}
\newcolumntype{R}{>{$}r<{$}}
\newcolumntype{F}{>{$}X<{$}}

In this section, we introduce the bounded semantics of HyperLTL, which
will be later used in Section~\ref{sec:qbf} to generate queries to a
QBF solver to aid solving the model checking problem.

\subsection{Bounded Semantics}
\label{subsec:bounded}

We assume the formula is closed and of the form:
$$
\quant_A \pi_A.\quant_B \pi_B\ldots\quant_Z \pi_Z.\psi
$$
where $\quant \in \{\forall, \exists\}$ and it has been converted into
negation-normal form (NNF) so that the negation symbol only appears in
front of atomic propositions, e.g., $\neg a_{\pi_A}$. Without loss of
generality and for the sake of clarity from other numerical indices,
we use roman alphabet as indices of trace variables. Thus, we assume that
$\Vars(\varphi) \subseteq  \{\pi_A, \pi_B, \dots, \pi_Z\}$.
The main idea of bounded model checking is to perform incremental
exploration of the state space of the systems by unrolling the systems
and the formula up-to a bound.
Let $k \geq 0$ be the unrolling {\em bound} and let
$\Tr=\tupleof{T_A\ldots T_Z}$ be a tuple of finite sets of finite
traces, one per trace variable.
We start by defining a satisfaction relation between HyperLTL
formulas for a bounded exploration $k$ and models $(\Tr,\Pi,i)$, where
$\Tr$ is the tuple of set of traces, $\Pi$ is a trace assignment
mapping (as defined in Section~\ref{sec:prelim}), and $i \in \zplus$ that points 
to the position of traces.
We will define different finite satisfaction relations for general
models (for $*=\textit{pes},\textit{opt},\textit{hpes},\textit{hopt}$):

\begin{itemize}

\item $\models_k^*$, the common satisfaction relation among all semantics, 
\vspace{.06cm}
\item $\Pmodels_k$, called {\em pessimistic} semantics,
\vspace{.06cm}
\item $\Omodels_k$, called {\em optimistic} semantics, and
\vspace{.06cm}
\item $\HPmodels_k$ and $\HOmodels_k$, variants of $\Pmodels_k$ and 
$\Omodels_k$, respectively, for Kripke structures that encode termination
  of traces (modeled as self-loops to provide infinite traces).
\end{itemize}

All these semantics coincide in the interpretation of quantifiers, Boolean
connectives, and in the interpretation of the temporal
operators up-to instant $k-1$, but differ in their assumptions about
unseen future events after the bound of observation $k$.

\paragraph{\textup{\textbf{Quantifiers.}}} The satisfaction relation for the
quantifiers is the following:

\noindent\begin{tabularx}{\textwidth}{RL@{\hspace{1em}}C@{\hspace{1em}}FR}\\[-0.3em]
    (\Tr, \Pi,0) \models_k^* & \exists \pi.\ \modified{\psi} & \text{iff} & \text{there is a } t \in 
 T_\pi: (\Tr,\Pi[\pi \rightarrow t],0) \models_k \psi, & (1)\\

 (\Tr, \Pi,0) \models_k^* & \forall \pi.\ \modified{\psi} & \text{iff} & \text{for all }\hspace{1.5em} t \in 
                                                               T_\pi: (\Tr,\Pi[\pi \rightarrow t],0) \models_k \psi. &(2)\\[0.6em]
\end{tabularx}
 
\paragraph{\textup{\textbf{Boolean operators.}}} For every $i\leq k$, we have:

\noindent\begin{tabularx}{\textwidth}{RL@{\hspace{1em}}C@{\hspace{1em}}FR}\\[-0.3em]
(\Tr, \Pi,i) \models_k^* & \tru & &  & (3) \\
(\Tr, \Pi,i) \models_k^* & a_\pi & \text{iff} & a \in \Pi(\pi)(i), & (4) \\
(\Tr, \Pi,i) \models_k^* & \neg a_\pi & \text{iff} & a \not\in \Pi(\pi)(i), & (5)\\
%
%
(\Tr, \Pi,i) \models_k^* & \psi_1 \vee \psi_2 & \text{iff} & (\Tr, \Pi,i) \models_k
\psi_1 \text{ or } (\Tr, \Pi,i) \models_k \psi_2, & (6) \\
(\Tr, \Pi,i) \models_k^* & \psi_1 \wedge \psi_2 & \text{iff} & (\Tr, \Pi,i) \models_k 
\psi_1 \text{ and } (\Tr, \Pi,i) \models_k \psi_2 & (7)\\[0.6em]
\end{tabularx}

\paragraph{\textup{\textbf{Temporal connectives.}}} The case where $(i<k)$ is common
between the optimistic and pessimistic semantics:

  \noindent\begin{tabularx}{\textwidth}{RL@{\hspace{1em}}C@{\hspace{1em}}FR}\\[-0.3em]
(\Tr, \Pi,i) \models_k^* & \X\psi & \text{iff} & (\Tr, \Pi,i+1) \models_k \psi & (8) \\
(\Tr, \Pi,i) \models_k^* & \psi_1 \U \psi_2 & \text{iff} & (\Tr, \Pi,i) \models_k 
\psi_2, \text{ or }\\
&&&\hspace{1em}(\Tr, \Pi,i) \models_k \psi_1 \text{ and } (\Tr, \Pi,i+1) \models_k \psi_1\U\psi_2 & (9)\\
(\Tr, \Pi,i) \models_k^* & \psi_1 \release \psi_2 & \text{iff} & (\Tr, \Pi,i) \models_k 
\psi_2 \text{, and }\\
&&&\hspace{1em} (\Tr, \Pi,i) \models_k \psi_1 \text{ or } (\Tr, \Pi,i+1) \models_k \psi_1\release\psi_2  & (10) \\[0.8em]
%
\end{tabularx}
  For $(i=k)$, in the pessimistic semantics the eventualities (including $\Next$)
  are assumed to never be fulfilled in the future, so the current instant $k$ is
  the last chance:
  
  \noindent\begin{tabularx}{\textwidth}{RL@{\hspace{1em}}C@{\hspace{1em}}FR}\\[-0.3em]
    (\Tr, \Pi,i) \Pmodels_k & \X\psi & \text{iff} & \text{never happens} & (P_1) \\
    (\Tr, \Pi,i) \Pmodels_k & \psi_1 \U \psi_2 & \text{iff} & (\Tr,\Pi,i)\Pmodels_k \psi_2 & (P_2) \\
   (\Tr, \Pi,i) \Pmodels_k & \psi_1 \release \psi_2 & \text{iff} & (\Tr,\Pi,i)\Pmodels_k\psi_1 \And \psi_2 & (P_3) \\[0.8em]
  \end{tabularx}

  \noindent On the other hand, in the optimistic semantics the
  eventualities are assumed to be fulfilled in the future:

    \noindent\begin{tabularx}{\textwidth}{RL@{\hspace{1em}}C@{\hspace{1em}}FR}\\[-0.3em]
    (\Tr, \Pi,i) \Omodels_k & \X\psi & \text{iff} & \text{always happens} & (O_1) \\
    (\Tr, \Pi,i) \Omodels_k & \psi_1 \U \psi_2 & \text{iff} & (\Tr,\Pi,i)\Omodels_k \psi_1 \Or \psi_2 & (O_2) \\
    (\Tr, \Pi,i) \Omodels_k & \psi_1 \release \psi_2 & \text{iff} & (\Tr,\Pi,i) \Omodels_k \psi_2 & (O_3) \\[0.8em]
  \end{tabularx}

  In order to capture the halting semantics, we assume that the Kripke
  structure is equipped with a predicate $\Halt$ that is true if the
  state corresponds to a halting state, and define the auxiliary
  predicate $\Halted\DefinedAs\AllHalt$ that holds whenever all traces
  have halted (and their final state will be repeated ad infinitum),
  where $\halt$ is an atomic proposition denoting the termination of a trace.
  Then, the halted semantics of the temporal case for $i=k$ in the
  pessimistic case consider the halting case to infer the actual value
  of the temporal operators on the (now fully known) trace:

    \noindent\begin{tabularx}{\textwidth}{RL@{\hspace{1em}}C@{\hspace{1em}}FR}\\[-0.3em]
      (\Tr, \Pi,i) \HPmodels_k & \X\psi & \text{iff} & (\Tr, \Pi,i)\models_k^*\Halted \text{ and } (\Tr, \Pi,i)\HPmodels_k \psi  & (\HP_1) \\
    (\Tr, \Pi,i) \HPmodels_k & \psi_1 \U \psi_2 & \text{iff} & (\Tr, \Pi,i)\HPmodels_k\psi_2 & (\HP_2) \\      
    (\Tr, \Pi,i) \HPmodels_k & \psi_1 \release \psi_2 & \text{iff} & (\Tr, \Pi,i)\HPmodels_k \psi_1 \And \psi_2, \text{ or } & \\
      &&& \hspace{0em} (\Tr,\Pi,i)\models_k^*\Halted \text{ and } (\Tr,\Pi,i)\HPmodels_k \psi_2  & (\HP_3) \\[0.8em]
  \end{tabularx}

\noindent{}Dually, in the halting optimistic case:

      \noindent\begin{tabularx}{\textwidth}{RL@{\hspace{0.8em}}C@{\hspace{0.8em}}FR}\\[-0.3em]
      (\Tr, \Pi,i) \HOmodels_k & \X\psi & \text{iff} & (\Tr, \Pi,i)\not\models_k^*\Halted \text{ or } (\Tr, \Pi,i)\HOmodels_k \psi  & (\HO_1) \\
      (\Tr, \Pi,i) \HOmodels_k & \psi_1 \U \psi_2 & \text{iff} & (\Tr,\Pi,i)\HOmodels_k \psi_2 \text{, or } & \\
      &&& \hspace{0em} (\Tr, \Pi,i)\not\models_k^*\Halted \text{ and } (\Tr, \Pi,i)\HOmodels_k \psi_1 & (\HO_2) \\
      (\Tr, \Pi,i) \HOmodels_k & \psi_1 \release \psi_2 & \text{iff} & (\Tr,\Pi,i)\HPmodels_k \psi_2  & (\HO_3) \\[0.8em]
  \end{tabularx}

\paragraph{\textup{\textbf{Complete semantics.}}} We are now ready to define the four semantics:
%

\noindent%
\begin{tabular}{rlll}
  $-$ & The pessimistic semantics $\Pmodels_k$ is comprised of rules &  $(1)$-$(10)$ & and $(P_1)$-$(P_3)$.\\
  $-$ & The optimistic semantics $\Omodels_k$ consists of rules & $(1)$-$(10)$ & and
        $(O_1)$-$(O_3)$.\\
  $-$ & The halting pessimistic semantics $\HPmodels_k$ use rules & $(1)$-$(10)$ & and $(\HP_1)$-$(\HP_3)$. \\
  $-$ & The halting optimistic semantics $\HOmodels_k$ use rules & $(1)$-$(10)$  & and
  $(\HO_1)$-$(\HO_3)$. \\
\end{tabular}


\subsection{The Logical Relation between Different Semantics}
  Observe that the pessimistic semantics is the semantics in the traditional BMC
  for LTL, where pending eventualities are considered to be
  unfulfilled.
  In the pessimistic semantics a formula is declared false unless it is
  witnessed to be true within the bound explored.
  In other words, formulas can only get ``truer'' with more information
  obtained by a longer unrolling.
%
  %
  Dually, the optimistic semantics considers a formula true unless
  there is evidence within the bounded exploration on the contrary.
  Therefore, formulas only get ``falser'' with further unrolling. For example,
  formula $\G p$ always evaluates to false in the pessimistic semantics. In
  the optimistic semantics, it evaluates to true upto bound $k$ if $p$ holds in
  all states of the trace upto and including $k$. However, if the formula
  evaluates to false at some point before $k$, then it evaluates to false for all $j \geq k$. 

  The following lemma formalizes this intuition in HyperLTL.

\begin{lemma}
  \label{lem:incr}
  Let $k\leq j$. Then,
\begin{enumerate}
  \item If $(\Tr,\Pi,0)\Pmodels_k\varphi$, then $(\Tr,\Pi,0)\Pmodels_j\varphi$.
  \item If $(\Tr,\Pi,0)\not\Omodels_k\varphi$, then $(\Tr,\Pi,0)\not\Omodels_j\varphi$.
  \item If $(\Tr,\Pi,0)\HPmodels_k\varphi$, then $(\Tr,\Pi,0)\HPmodels_j\varphi$.
  \item If $(\Tr,\Pi,0)\not\HOmodels_k\varphi$, then $(\Tr,\Pi,0)\not\HOmodels_j\varphi$.
\end{enumerate}
\end{lemma}

In turn, the verdict obtained from the exploration up-to $k$ can (in
some cases) be used to infer the verdict of the model checking
problem.
As in classical BMC, if the pessimistic semantics find a model, then
it is indeed a model.
Similarly, if the optimistic semantics fail to find a model, then
there is no model.
The next lemma formally captures this intuition.

\begin{lemma}[Infinite inference]
  \label{lem:infinite-inference}
  The following hold for every $k$,
  \begin{enumerate}
  \item If $(\Tr,\Pi,0)\Pmodels_k\varphi$, then
    $(\Tr,\Pi,0)\models\varphi$.
  \item If $(\Tr,\Pi,0)\not\Omodels_k\varphi$, then
    $(\Tr,\Pi,0)\not\models\varphi$.
  \item If $(\Tr,\Pi,0)\HPmodels_k\varphi$, then
    $(\Tr,\Pi,0)\models\varphi$.
  \item If $(\Tr,\Pi,0)\not\HOmodels_k\varphi$, then
    $(\Tr,\Pi,0)\not\models\varphi$.
    \end{enumerate}
\end{lemma}

%% file: examples.tex
\subsection{Examples}
\label{sec:examples-bounded}

%
Consider the Kripke structure in Fig.~\ref{fig:kripke}, bound $k=3$, and formula
$$
\varphi_1 = \forall \pi_A. \exists \pi_B.  \big((p_{\pi_A} \not\leftrightarrow p_{\pi_B}) \release \neg q_{\pi_A} \big)
$$
It is easy to see that instantiating $\pi_A$ with trace
$s_0 s_1 s_2 s_4$ is a trace $\pi_A$ of the negation, $\neg\varphi_1$ as follows, in the pessimistic
semantics.
$$
\neg\varphi_1 = \exists \pi_A. \forall \pi_B.  \big((a_{\pi_A} \leftrightarrow p_{\pi_B}) \until q_{\pi_A} \big)
$$
By Lemma~\ref{lem:infinite-inference}, this counterexample shows that
the kripke structure is a model of $\neg\varphi_1$ in the infinite
semantics as well.
That is, $\krip \Pmodels_3 \neg\varphi_1$ and, hence,
$\krip \models \neg \varphi_1$, so $\krip \not\models \varphi_1$ .%


%
%
Consider again the same Kripke structure, bound $k=3$, and formula
\[
\varphi_2 = \forall \pi_A. \exists \pi_B.  \F (p_{\pi_A} \leftrightarrow q_{\pi_B})
\]
To disprove $\varphi_2$, we need to find a trace $\pi_A$ such that for all
other $\pi_B$, proposition $q$ in $\pi_{B}$ always disagrees with $p$ in $\pi_A$,
as the following formula,
\[
\neg\varphi_2 = \exists \pi_A. \forall \pi_B.  \G (p_{\pi_A} \not\leftrightarrow q_{\pi_B})
\]
It is straightforward to observe that such a trace $\pi_A$ does not exist.
By Lemma~\ref{lem:infinite-inference}, proving the formula is not satisfiable 
upto bound 3 in the optimistic semantics implies that $\krip$ is not a model of 
$\neg\varphi_2$ in the infinite semantics. 
That is, $\krip \not\Omodels_3 \neg\varphi_2$ implies
$\krip \not\models \neg\varphi_2$. 
Hence, we conclude $\krip \models \varphi_2$.
%

%
%
%

Consider again the same Kripke structure which has two terminating states,
$s_3$ and $s_4$, labeled by atomic proposition $\halt$ with only a
self-loop. Let $k=3$, and formula,
\[
  \varphi_3 = \forall \pi_A. \exists \pi_B.  (\neg q_{\pi_B} \until \neg p_{\pi_A})   
\]
To disprove, we want to find a trace $\pi_A$ that fulfills the negation, 
\[
  \neg \varphi_3 = \exists \pi_A. \forall \pi_B.  (q_{\pi_B} \release p_{\pi_A})
\]
Take the halting state in to consideration, $s_0 s_1 s_3 $ 
is a trace of the form $\{p\}^\omega$.
It satisfies the halting optimistic semantic of $\release$ in $s_3$ because 
of the  halting condition.
By Lemma~\ref{lem:infinite-inference}, the fulfillment of formula implies that in infinite semantics 
it will be fulfilled as well. 
That is, $\krip \HPmodels_3 \neg\varphi_3$ implies
$\krip \models \neg\varphi_3$. Hence, $\krip \not\models \varphi_3$.        

Consider again the same Kripke structure with halting states and formula,
\[
  \varphi_4 = \forall \pi_A. \exists \pi_B.  \F \always (p_{\pi_A} \not\leftrightarrow p_{\pi_B})
\]
A counterexample is an instantiation of $\pi_A$ such that for all
$\pi_B$, both traces will always eventually agree on $p$ as follows,
\[
  \neg\varphi_4 = \exists \pi_A. \forall \pi_B.  \always \F (p_{\pi_A} \leftrightarrow p_{\pi_B})
\]
Trace $s_0 s_1 s_2 s_4$ is of the form $\{p\}\{p\}\{p\}\{r, \halt\}^\omega$
with $k=3$.
This trace never agrees with a trace that ends in state $s_3$ (which
is of the form $\{p\}^\omega$) and vice versa.
%
%
By Lemma~\ref{lem:infinite-inference}, the absence of counterexample 
upto bound 3 in the halting optimistic semantics implies that 
$\krip$ is not a model of $\neg\varphi_4$ in the infinite semantics.
That is, $\krip \not\HOmodels_3 \neg\varphi_4$ implies
$\krip \not\models \neg\varphi_4$.
Hence, we conclude $\krip \models \varphi_4$.


%% file: qbf.tex
\section{Reducing BMC to QBF Solving}
\label{sec:qbf}

We describe in this section (1) how to generate a QBF query from an
instance of the model checking problem, and (2) what can be inferred
in each case from the outcome of the QBF solver about the model
checking problem.

\subsection{QBF-based Solution}

Given a family of Kripke structures $\Kr$, a HyperLTL formula
$\varphi$, and bound $k \geq 0$, our goal is to construct a quantified
Boolean formula $ \qbf{\Kr, \varphi}_k$ whose satisfiability can be
used to infer whether or not $\Kr \models \varphi$.
We first describe how to encode the model and the formula, and then
how to combine the two to generate the QBF query.

\paragraph{Encoding the models.} The unrolling of the transition
relation of a Kripke structure $\krip_A = \ktuple$ up to bound $k$ is
analogous to the BMC encoding for LTL~\cite{cbrz01}.
First, note that the state space $S$ can be encoded with a
(logarithmic) number of bits in $|S|$.
We introduce additional variables $n_0,n_1,\ldots$ to encode the state
of the Kripke structure and use $\AP^*=\AP \cup \{n_0,n_1,\ldots\}$
for the extended alphabet that includes the encoding $S$.
In this manner, the set of initial states of a Kripke structure is a
Boolean formula over $\AP^*$.
For example, for the Kripke structure $K_A$ in Fig.~\ref{fig:kripke}
the set of initial states (in this case $S_\init=\{s_0\}$) corresponds
to the following Boolean formula:
\[
  I_A := (\neg {n_0} \land \neg {n_1} \land \neg {n_2}) \land p \land
  \neg q \land \neg \halt
\]
assuming that $(\neg {n_0} \land \neg {n_1} \land \neg {n_2})$ encodes
state $s_0$ (we need three bits to encode five states.)
Similarly $R_A$ is a binary relation that encodes the transition
relation $\delta$ of $K_A$ (encoding the relation between a state and
its successor).
The encoding into QBF works by introducing fresh Boolean variables (a
new copy of $\AP^*$ for each Kripke structure $\krip_A$ and position),
and then producing a Boolean formula that encodes the unrolling up-to
$k$.
We use $x_A^i$ for the set of fresh copies of the variables $\AP^*$ of
$K_A$ corresponding to position $i\in[0,k]$.
Therefore, there are $k|x_A| = k |\AP_A^*|$ Boolean variables to represent the
unrolling of $K_A$.
We use $I_A(x)$ for the Boolean formula (using variables from $x$)
that encodes the initial states, and $R_A(x,x')$ (for two copies of the
variables $x$ and $x'$) for the Boolean formula whether $x'$ encodes a
successor states of $x$.

%
%

For example, for bound $k=3$, we unroll the transition relation up-to
$3$ as follows,
\[
\qbf{\krip_A}_3 = I_A(x^0_A) \land R_A(x_A^0, x_A^1) \land R(x_A^1, x_A^2) \land R(x^2_A, x_A^3)
\]
which is the Boolean formula representing valid traces of length $4$,
using four copies of the variables $\AP^*_A$ that represent the Kripke
structure $K_A$.
 
\paragraph{Encoding the inner LTL formula.} The idea of the construction of the
inner LTL formula is analogous to the standard BMC as well, except for the
choice of different semantics described in Section~\ref{sec:bmc}.
In particular, we introduce the following inductive construction and
define four different unrollings for a given $k$:
$\QBFpes{i}{k}{\cdot}$, $\QBFopt{i}{k}{\cdot}$,
$\QBFhpes{i}{k}{\cdot}$, and $\QBFhopt{i}{k}{\cdot}$.

\begin{itemize}
\item \textbf{Inductive Case}: Since the semantics only differ on the
  temporal operators at the end of the unrolling, the inductive case
  is common to all unrollings and we use $\QBFany{i}{k}{\cdot}$ to
  mean any of the choices of semantic (for
  $*=\textit{pes},\textit{opt},\textit{hpes},\textit{hopt}$).
  For all $i \leq k$:
\[
\begin{array}{l@{\hspace{1em}}c@{\hspace{1em}}l}
\QBFany{k}{i}{p_\pi} & := & p_\pi^i \\
\QBFany{k}{i}{\neg p_\pi} & := & \neg p_\pi^i \\
 %
%
\QBFany{k}{i}{\psi_1\OR\psi_2} & := & \QBFany{k}{i}{\psi_1} \OR\QBFany{k}{i}{\psi_2}\\
\QBFany{k}{i}{\psi_1\AND\psi_2} & := & \QBFany{k}{i}{\psi_1} \AND \QBFany{k}{i}{\psi_2}\\

\QBFany{k}{i}{\psi_1 \, \until \, \psi_2} & := & \QBFany{k}{i}{\psi_2} \; \vee \; 
\Big(\QBFany{k}{i}{\psi_1} \, \wedge \, \QBFany{k}{i+1}{\psi_1 \, \until 
\, \psi_2}\Big)\\
\QBFany{k}{i}{\psi_1 \, \release \, \psi_2} & := & \QBFany{k}{i}{\psi_2}  \;\AND\;
\Big(\QBFany{k}{i}{\psi_1} \, \OR \, \QBFany{k}{i+1}{\psi_1 \, \release
\, \psi_2}\Big)\\
\QBFany{k}{i}{\X\psi} & := & \QBFany{k}{i+1}{\psi}\\
\end{array}
\]
Note that, for a given path variable $\pi_A$, the atom $p^i_{\pi_A}$
that results from $\QBFany{k}{i}{p_\pi}$ is one of the Boolean
variables in $x^i_A$.

\item For the \textbf{base case}, the formula generate is different
  depending on the intended semantics:
  \[
    \begin{array}{rcl@{\hspace{3em}}rcl}
      \QBFpes{k}{k+1}{\psi} & := & \fals & \QBFopt{k}{k+1}{\psi} &:= & \tru\\
      \QBFhpes{k}{k+1}{\psi} & := & \QBFhpes{k}{k}{\Halted\,} \And \QBFhpes{k}{k}{\psi} & \QBFhopt{k}{k+1}{\psi} & := & \QBFhopt{k}{k}{\Halted\,} \rightarrow \QBFhopt{k}{k}{\psi}
    \end{array}
  \]
  Note that the base case defines the value to be assumed for the
  formula after the end $k$ of the unrolling, which is spawned in the
  temporal operators in the inductive case at $k$.
  The pessimistic semantics assume the formula to be false, and the
  optimistic semantics assume the formula to be true.
  The halting cases consider the case at which the traces have halted
  (using in this case the evaluation at $k$) and using the unhalting
  choice otherwise.
\end{itemize}

\paragraph{Combining the encodings.}
Now, let $\varphi$ be a HyperLTL formula of the form
$\varphi = \quant_A\pi_A.\quant_B\pi_B.\dots.\quant_Z\pi_Z.\psi$ and
$\Kr = \langle \krip_A, \krip_B, \dots, \krip_{Z} \rangle$.
Combining all the components, the encoding of the HyperLTL BMC problem
in QBF is the following (for
$*=\textit{pes},\textit{opt},\textit{hpes},\textit{hopt}$):
\[
\qbf{\Kr, \varphi}^*_k = \quant_A\overline{x_A}.\quant_B\overline{x_B}\cdots.\quant_Z\overline{x_Z} \Big( \qbf{\krip_A}_k 
\circ_A \qbf{\krip_B}_k \circ_B \cdots \qbf{\krip_Z}_k \circ_Z 
\QBFany{0}{k}{\psi} \Big)
\]
\noindent{}where $\QBFany{0}{k}{\psi}$ is the choice of semantics and,
 $\circ_j = \wedge$ if $\quant_j = \exists$ and
$\circ_j = \rightarrow$ if $\quant_j = \forall$, for $j \in \Vars(\varphi)$.

\paragraph{Example.} Consider formula $\varphi_1$ in Section~\ref{sec:examples-bounded}, whose negation is the following:
$$
\neg\varphi_1 := \exists \pi_A. \forall \pi_B. \underbrace{\big((p_{\pi_A} \leftrightarrow p_{\pi_B}) \until q_{\pi_A} \big)}_{\neg \psi}
$$
The unrolling of $\neg\psi$ using the pessimistic semantics is

\begin{align*}
  \qbf{\neg\psi}^{\PES}_{0,3} &= \qbf{\big((p_{\pi_A} \leftrightarrow p_{\pi_B}) \until q_{\pi_A} \big)}^{\PES}_{0,3} = \\
  & =  q_{\pi_A}^0 \lor 
  	\Big(({p_{\pi_A}^0} \leftrightarrow {p^0_{\pi_B}})  
	\land  \Big( {q^1_{\pi_A}} \lor 
			\Big( ({p^1_{\pi_A}} \leftrightarrow {p^1_{\pi_B}}) 
		 \land \Big({q^2_{\pi_A}} \lor 
		 \Big(({p^2_{\pi_A}} \leftrightarrow {p^2_{\pi_B}})
		 \land \Big(q^3_{\pi_A}\Big)\Big)\Big)\Big)\Big)\Big) \\
\end{align*}

Note that in the final encoding, for example the collection $x_A^2$,
contains all variables of $\AP^*$ of $K_A$ (for example, $p_{\pi_A}^2$)
connecting to the corresponding valuation for $p_{\pi_A}$ in the trace
of $K_A$ at step $2$ in the unrolling of $K_A$.
In other words, the formula $\qbf{\neg\psi}^{\PES}_{0,3}$ uses
variables from $x_A^0,x_A^1,x_A^2,x_A^3$ and $x_B^0,x_B^1,x_B^2,x_B^3$
(that is, from $\overline{x_A}$ and $\overline{x_B}$).
To combine the model description with the encoding of the HyperLTL
formula, we use two identical copies of the given Kripke structure to
represent different paths $\pi_A$ and $\pi_B$ on the model, denoted as
$\krip_A$ and $\krip_B$.
The resulting formula is:
\[
\begin{array}{c c l}
\qbf{\Kr, \neg\varphi}_3 & := & \exists \overline{x_A}.\forall \overline{x_B}.\big( \qbf{\krip_A}_3 \AND  (\qbf{\krip_B}_3 \rightarrow \qbf{\neg\varphi}_{0,3}^\PES ) \big) \\[0.6ex]
\end{array}
\]  
The sequence of assignment
$\{(\neg n_2, \neg n_1, \neg n_0, p, \neg q)^0$,
$(\neg n_2, \neg n_1, n_0, p, \neg q)^1$,\linebreak
$(\neg n_2, n_1, \neg n_0, p, \neg q)^2$,
$( n_2, \neg n_1, \neg n_0, \neg p, q)^3\}$ on $\krip_A$,
corresponding to the trace $s_0 s_1 s_2 s_4$, satisfies
$\qbf{\neg\varphi}_{0,3}^\PES$ for all traces on $\krip_B$.
The satisfaction results shows that $\qbf{\Kr, \neg\varphi}^\PES_3$ is
true, indicating that a witness of violation is found.
Theorem~\ref{thm:qbf-mc}, by a successful detection of a
counterexample witness, and the use of the pessimistic semantics,
allows to conclude that $\Kr\not\models\varphi$.\qed
%

%
%
%
%
%
%
%
%

\subsection{Soundness Results}

\begin{lemma}
  \label{lem:qbf-bmc}
  Let $\varphi$ be a closed HyperLTL formula and $\Tr=\Traces(\Kr)$ be
  an interpretation.  For
  $*=\textit{pes},\textit{opt},\textit{hpes},\textit{hopt}$, it holds
  that
  \[ 
    \qbf{\Kr,\varphi}_k^{*} \text{ is satisfiable if and only if }
    (\Tr,\Pi_{\emptyset},0)\models^*_k\varphi.
  \]
\end{lemma}

\begin{proof}[sketch]
  The proof proceeds in two steps.  First, let $\psi$ be the largest
  quantifier-free sub-formula of $\varphi$. Then, every tuple of
  traces of length $k$ (one for each $\pi$) is in one to one
  correspondence with the collection of variables $p_\pi^i$, that
  satisfies that the tuple is a model of $\psi$ (in the choice
  semantics) if and only if the corresponding assigment makes
  $\qbf{\psi}^*_0$.
  Then, the second part shows inductively in the stack of quantifiers
  that each subformula obtained by adding a quantifier is satisfiable
  if and only the semantics hold. \qed
\end{proof}
  
Lemma~\ref{lem:qbf-bmc}, together with
Lemma~\ref{lem:infinite-inference}, allows to infer the outcome of the
model checking problem from satisfying (or unsatisfying) instances of
QBF queries, summarized in the following theorem.

\begin{theorem}
  \label{thm:qbf-mc}
  Let $\varphi$ be a HyperLTL formula. Then,
  \begin{compactenum}
  \item For $*=\PES,\HPES$, if
    $\qbf{\Kr,\neg\varphi}_k^{*}$ is satisfiable then
    \hspace{1.171em}$\Kr\not\models\varphi$.
  \item For $*=\OPT,\HOPT$, if
    $\qbf{\Kr,\neg\varphi}_k^{*}$ is unsatisfiable then
    $\Kr\models\varphi$.
  \end{compactenum}
\end{theorem}

\paragraph{Example.} Finally, we make the connection between satisfiability of QBF and
the infinite semantics of the examples in Section~\ref{sec:examples-bounded}
using Theorem~\ref{thm:qbf-mc}.
Table~\ref{tab:example} illustrates what the different semantics
allows to soundly conclude.

\renewcommand{\arraystretch}{1.3}

\begin{table}[h]
  \centering 
  \scalebox{.78}{
  \begin{tabular}{K{1.5cm}  K{1.1cm} || K{4cm} |  K{4cm} | K{4cm} |} 
    \cline{3-5}
    & & \multicolumn{3}{|c|}{\bf Semantics}\\
    \hline
    \multicolumn{1}{|c||}{\bf Formula}  & {\bf Bound} & {\em pessimistic} &   {\em optimistic} & {\em halting}  \\ [0.5ex] 
\hline\hline 
\multicolumn{1}{ |c||  }{\multirow{2}{*} {$\varphi_1$}} & $k=2$ & UNSAT (inconclusive) & SAT (inconclusive)  & UNSAT (inconclusive) \\
\multicolumn{1}{ |c||  }{}                       
& $k=3$ & SAT (\textbf{\em counterexample}) & SAT (inconclusive) & UNSAT (inconclusive)\\
\hline\hline
\multicolumn{1}{ |c||  }{\multirow{2}{*} {$\varphi_2$}} 	& $k=2$ & UNSAT (inconclusive) & SAT (inconclusive) & UNSAT (inconclusive) \\ 
\multicolumn{1}{ |c||  }{}					 	& $k=3$ & UNSAT (inconclusive) & UNSAT (\textbf{\em proved}) & UNSAT (inconclusive)\\
\hline\hline
\multicolumn{1}{ |c||  }{\multirow{2}{*} {$\varphi_3$}} 	& $k=2$ & UNSAT (inconclusive) & UNSAT (inconclusive) & non-halted (inconclusive) \\ 
\multicolumn{1}{ |c||  }{}					 	& $k=3$ & UNSAT (inconclusive) & UNSAT (inconclusive) & halted (\textbf{\em counterexample}) \\
\hline\hline
\multicolumn{1}{ |c||  }{\multirow{2}{*} {$\varphi_4$}} 	& $k=2$ & UNSAT (inconclusive) & UNSAT (inconclusive) & non-halted (inconclusive) \\ 
\multicolumn{1}{ |c||  }{}					 	& $k=3$ & UNSAT (inconclusive) & UNSAT (inconclusive) & halted (\textbf{\em proved}) \\
\hline
  \end{tabular}
  }
  \caption{Comparison of Properties with Different Semantics} 
  \label{tab:example}
\end{table}

\renewcommand{\arraystretch}{1}

\input{qbf_example}


%% file: qbf_example.tex
%


%% file: cases.tex
\section{Descriptions of Case Studies}
\label{sec:cases}

\input{algs} 

In this section, we introduce a rich set of case studies to verify and
falsify hyperproperties for different systems.
These include proving symmetry of the Bakery Algorithm mutual
exclusion protocol, linearizability of the SNARK algorithm,
non-interference in multi-threaded programs and fairness in
non-repudiation protocols.~\cite{cfst19, ddgflmmss04, sv98, jmm11}
We also show to strategies can be synthesized for robotic planning and
mutation testing using our QBF encoding~\cite{wnp19, ftw19}.
%

\subsection{Case study 1: Symmetry in the Bakery Algorithm}
We first investigate the symmetry property in Lamport's Bakery
algorithm for enforcing mutual exclusion in a concurrent program.~\cite{cfst19}
The Bakery algorithm works as follows.
When a process $p$ intends to enter the critical section, $p$ draws a
``ticket'' modeled by a number.
When more than one process attempt to enter the critical section, the
process with the smallest ticket number enters first, while other
processes wait.
In a concurrent program, it is also possible that two or more
processes hold tickets with same number if they drew tickets
simultaneously.
To solve this tie, when processes with the same ticket try to access
the critical section, the process with smaller process ID enters first
while the other processes wait.
The Bakery algorithm is shown in Algorithm~\ref{alg:bbb}.

\AlgBakery

We are interested in studying the symmetry property, which informally
states that no specific process has special privileges in terms of a
faster access to the critical section.
We use the atomic proposition $\select$ to represent the process
selected to proceed in the next state, and $\pause$ to indicate if the
processes are both not moving.
Each process $P_n$ has a program counter denoted by $\pc(P_n)$.
The symmetry property for the Bakery algorithm is formally express as
follows.
For all traces $\pi_{A}$, there exists a trace $\pi_{B}$, such that if both
traces at every step select the next process to execute symmetrically,
then the program counter of each process would be completely symmetric
as well.
For example, consider two processes $P_0$ and $P_1$ and let trace
$\pi_{A}$ select $P_0$ \; iff \; trace $\pi_{B}$ selects $P_1$, and $\pi_{A}$
select $P_1$ \; iff \; $\pi_{B}$ selects $P_0$.
Such a dual choice of selection is presented as
$\sym(\select_{\pi_{A}}, \select_{\pi_{B}})$.
We are ready to describe the symmetry property as the following
HyperLTL formula:

\begin{center}
\centering 
\begin{tabular}{ |c|c| } 
\hline
\thead{Symmetry} & \thead{ $\varphi_{\sym} = \forall \pi_{A}. \exists\pi_{B}.\ 
\always\Big(\sym(\select_{\pi_{A}}, \select_{\pi_{B}}) \land (\pause_{\pi_{A}}= 
\pause_{\pi_{B}}) \; \land  $ \\ $\hfill  \big(\pc(P_0)_{\pi_{A}} = 
\pc(P_1)_{\pi_{B}}\big) \land \big(\pc(P_1)_{\pi_{A}} = \pc(P_0)_{\pi_{B}}\big)\Big)$} \\
\hline
\end{tabular}
\end{center}




\subsection{Case study 2: Linearizability of the SNARK Algorithm}
Next, we investigate whether the SNARK algorithm~\cite{ddgflmmss04} satisfies the
linearizability property.

\AlgSNARK

Linearizability is a correctness property of concurrent libraries or
datatypes~\cite{hw90}.
The $\history$ of the execution of a concurrent datatype, is the
sequence of method {\em invocations} by the different threads and the
{\em response} observed.
A $\history$ is {\em linearizable}, if there exists a
sequential order of invocations and responses, such that the same
responses could be produced with atomic executions of the methods
invoked.
A concurrent datatype is linearizable if all possible histories are
linearizable.
In~\cite{bss18}, the authors show that linearizability is a
hyperproperty of the form $\forall\exists$, where the domain of the
universal quantifier ranges over all possible executions of the
concurrent data structure and the domain of the existential quantifier
ranges over all possible executions of a sequential implementation of
the data structure (or over the sequential reference implementation or
declarative specification of the datatype).
Thus, reasoning about linearizability requires our multi-model
semantics introduced in Section~\ref{sec:prelim}.

The SNARK algorithm~\cite{ddgflmmss04} is a concurrent implementation
of a double-ended queue data structure (the pseudo-code is shown in
Algorithm~\ref{alg:snark}).
It uses double-compare-and-swap (DCAS) with doubly linked-list that stores 
values in nodes while each node is connected to its two neighbors, $L$ and $R$. 
When a modification of data happens, e.g., by invoking
\code{pushRight()} or \code{popLeft()}, SNARK performs a DCAS by
comparing two memory locations to decide if such modification is
appropriate.

We define linearizability as a hyperproperty using two different
models.
Let $\pi_A$ denote the trace variable over the traces of the {\em
  concurrent program} (in this case SNARK).
This program is created by allows multiple to execute each method with
interleavings.
Let $\pi_B$ represents the trace variable over traces of the sequential 
implementation of a double-ended queue (i.e., the specification), where only 
atomic invocations are allowed. 
The HyperLTL formula that specifies linearizability is:

\begin{center}
\centering 
\begin{tabular}{ |c|c| } 
\hline
\thead{Linearizability} & \thead{ $\varphi_{\linearizability} = \forall  
\pi_{A}. \exists \pi_{B}. \ {\always}(\history_{\pi_A} \,
\leftrightarrow \, \history_{\pi_B})$} \\
\hline
\end{tabular}
\end{center}



\subsection{Case study 3: Non-interference in Typed Multi-threaded Programs}
We also investigate {\em non-interference} in a multi-threaded 
program with type system.
Non-interference is a security policy that states that low-security
variables are independent from the high-security variables, thus,
preserving secure information flow.
Each variable is labeled as a {\em high-variable} (high security) or {
  \em low-variable} (low security).
Non-interference requires that all information about a high-variable
cannot be inferred by observing any the values of a low-variable.
In this case study, we look at a concurrent system example
from~\cite{sv98}, which contains three threads $\alpha$, $\beta$, and
$\gamma$.
The variables are assigned with different security level as follows:
{\em PIN}, {\em trigger0}, and {\em trigger1} are as high-variables,
and {\em maintrigger}, {\em mask}, and {\em result} are low-variables.

\AlgMultiThreaded

Assuming that thread scheduling is fair, the program satisfies
non-interference, if for all executions, there exists another
execution that starts from a different high-inputs (i.e., the values
of {\em PIN} are not equal) and at termination point, they are in
low-equivalent states (i.e., the values of {\em Result} are equal).
Furthermore, in order to search for a witness of non-interference
violation in bounded time, we also consider {\em halting} as
introduced in Section~\ref{sec:boundedHyperLTL}.
In this particular program, the execution terminates when the low-variable {\em 
MASK} contains value zero. 
The corresponding HyperLTL formula is:

\begin{center}
\centering 
\begin{tabular}{ |c|c| } 
\hline
\thead{$\NI$} & \thead{ $\varphi_{\NI} =\forall \pi_{A}. \exists \pi_{B}. 
\big(\PIN_{\pi_{A}} 
\neq \PIN_{\pi_{B}}\big) \land \Big((\neg \halt_{\pi_{A}} \lor \neg 
\halt_{\pi_{B}})$ \\ 
$\hfill \mathcal U \ \big((\halt_{\pi_{A}} \land \halt_{\pi_{B}}) \land 
(\Result_{\pi_{A}} = \Result_{\pi_{B}}) \big) \Big)\ $} \\
\hline
\end{tabular}
\end{center}
where atomic proposition $\halt$ denotes the halting state ({\em 
MASK} contains a zero bit) and by abuse of notation $\PIN_\pi$ (respectively, 
$\Result_\pi$) denotes the value of $\PIN$ (respectively, $\Result$) in trace 
$\pi$.
%


\subsection{Case study 4: Fairness in Non-repudiation Protocols}
A non-repudiation protocol consists of three parties: a message sender
($\partyA$), a message receiver ($\partyB$), and a {\em trusted third
  party} $T$.
In a message exchange event, the message receiver should obtain a
receipt from the sender, named {\em non-repudiation\ of\ origin}
($\NRO$), and the message sender should end up having an evidence
named {\em non-repudiation of receipt} ($\NRR$).
The three participants can take the following actions: 
\begin{align*}
\Act_P & = \{ P \rightarrow Q:m,\  P \rightarrow T:m ,\ P \rightarrow Q:\NRO , P \rightarrow T:\NRO ,\  P:\skipp \}\\
\Act_Q & = \{ Q \rightarrow P:\NRR ,\  Q \rightarrow T:\NRR ,\ Q:\skipp \}\\
\Act_T & = \{ T \rightarrow P:\NRR ,\  T \rightarrow Q:\NRO ,\ T:\skipp \}
\end{align*}

In this case study, we evaluate two different models of trusted third party 
from~\cite{jmm11}. 
First, we pick an incorrect implementation from~\cite{jmm11}, named
$T_{\incorrect}$, which $\partyB$ can choose not to send out $\NRR$
after receiving $\NRO$.
We also consider a correct implementation of the protocol.
Both versions are show in Alg.~\ref{alg:non-rep}.

\AlgNonRepudiation

A {\em fair} non-repudiation protocol guarantees that two parties can
exchange messages fairly without any party being able to deny sending
out evidence while having received an evidence.
Furthermore, we say that a trace is {\em effective} if {\em message},
$\NRR$, and $\NRO$ are all received.
Assuming that each party will take turns and take different actions,
the fairness of non-repudiation protocol can be defined as a
hyperproperty as follows.
There exists an {\em effective} trace $\pi_{A}$, such that for all other
traces $\pi_{B}$, if $\partyA$ in both traces always take the same action
while $\partyB$ behave arbitrarily, or both $\partyB$ take the same
action and $\partyA$ behave arbitrarily, then for $\pi_{B}$, eventually
$\NRR$ gets received by $\partyA$ if and only if $\NRO$ gets received
by $\partyB$.

The complete specification for non-repudiation is the following:
%

\begin{center}
\centering 
\begin{tabular}{ |c|c| } 
\hline
\thead{Fairness} & \thead[l]{ 
$\varphi_{\fair} = \exists \pi_{A}. \forall \pi_{B}.$  $({\F} m_{\pi_{A}}) \land ({\F} 
\NRR_{\pi_{A}}) \land ({\F} \NRO_{\pi_{A}}) \; \land $\\
$\hspace{1cm} \Big(({\always} \bigwedge_{act \in \Act_P} act_{\pi_{A}} \leftrightarrow 
act_{\pi_{B}} )\ \rightarrow \big(({\F} \NRR_{\pi_{B}}) \leftrightarrow\ ({\F} 
\NRO_{\pi_{B}})\big)\Big) \;  \land $\\
$\hspace{1cm}\Big(({\always} \bigwedge_{act \in \Act_Q} act_{\pi_{A}} \leftrightarrow 
act_{\pi_{B}} )\ \rightarrow \big(({\F} \NRR_{\pi_{B}}) \leftrightarrow\ ({\F} 
\NRO_{\pi_{B}})\big)\Big)$}\\
\hline
\end{tabular}
\end{center}
Observe that trace $\pi_{A}$ expresses effectiveness (i.e., an honest behavior 
of all parties), while trace $\pi_{B}$ is a trace that behaves similarly to 
trace $\pi_{A}$ as far as the actions of $\partyA$ or $\partyB$ are concerned 
while ensuring fair receipt of $\NRR$ and $\NRO$. 



\subsection{Case study 5: Privacy-Preserving Path Planning for Robots}
In addition to model checking problems, inspired by the work
in~\cite{wnp19}, we explore other applications of our QBF encoding
that also involve hyperproperties with quantifier alternation.
One such application is searching the optimal solution for robotic
planning.
For example, given a 2-D grid with an initial state and a goal state,
a shortest path from initial state to goal state is a trace $\pi_{A}$,
such that $\pi_{A}$ reaches the goal state and for all other traces
$\pi_{B}$, $\pi_{B}$ has not reached the goal state before $\pi_{A}$ has.
In other words, the shortest path is a path on the grid that reaches
the goal state before all other paths.
We express this specification as the following hyperproperty:

\begin{center}
\centering 
\begin{tabular}{ |c|c| } 
\hline
\thead{Shortest Path} & \thead{ $\varphi_{\mathit{sp}} = \exists \pi_{A}. \forall 
\pi_{B}. ( \neg \goal_{\pi_{B}}\ \mathcal U\  \goal_{\pi_{A}} )$} \\
\hline
\end{tabular}
\end{center}
where the atomic proposition $\goal$ denotes that the path has reached the goal 
state.

To further analyze the result, we also consider that traces halt.
An optimal path searching should terminate when the shortest path is found 
because when a shortest path has been discovered on the map, any further 
exploration will not affect the outcome .

Besides optimal solution searching, HyperLTL also allows us to specify the 
{\em robustness} of paths that are derived by uncertainty in robotic planning. 
For example, instead of one single initial state, we now consider a map with a 
set of initial states. 
We are interested in a strategy that can help all traces to reach the
goal state regardless of which initial state the path start from.
The robust strategy searching problem can be presented as follows.
There exists a robust path $\pi_{A}$, such that for all paths $\pi_{B}$
starting from as arbitrary state from the set of initial states,
$\pi_{B}$ is able to reach the goal state using the same strategy as
$\pi_{A}$.
We use the proposition $\strategy$ to represent the sequence of
movements the path takes.
We write the formula as follows:

\begin{center}
\centering 
\begin{tabular}{ |c|c| } 
\hline
\thead{Robustness} & \thead{ $\varphi_{\mathit{rb}} = \exists \pi_{A}. \forall 
\pi_{B}.\  (\strategy_{\pi_{B}} \leftrightarrow \strategy_{\pi_{A}}) \until (\goal_{\pi_{A}} \land \goal_{\pi_{B}})$} \\
\hline
\end{tabular}
\end{center}

\subsection{Case study 6: Generate Mutants in Mutation Testing}
Another application of hyperproperty with quantifier alternation is
the efficient generation of test suites for mutation testing.
We look at the beverage machine model from~\cite{ftw19}.
The beverage machine has three possible inputs: {\em request}, {\em
  fill}, or {\em none}.
Based on the input, the machine may output {\em coffee}, {\em tea}, or
{\em none}.
We also use an atomic proposition $\mut$ to mark mutated traces, and
$\neg\mut$ for non-mutated traces.
In this non-deterministic model, a potentially killable mutant is a
trace (mutated) trace $\pi_{A}$ such that, for all other (non-mutated)
$\pi_{B}$, if they have same inputs as $\pi_{A}$, then the outputs eventually
diverge.

\begin{center}
\centering 
\begin{tabular}{ |c|c| } 
\hline
\thead{Mutant in \\ Non-det Model} & \thead{ $\exists \pi_{A} \forall \pi_{B} (\mut_{\pi_{A}} \land \neg \mut_{\pi_{B}}) \land
 \big((\inputt_{\pi_{A}} \leftrightarrow \inputt_{\pi_{B}})\ \mathcal U\ (\outputt_{\pi_{A}} \not\leftrightarrow \outputt_{\pi_{B}})\big)$} \\
\hline
\end{tabular}
\end{center}


%% file: algs.tex
\newcommand{\AlgBakery}{
\begin{algorithm}[ht]
\caption{Bakery}
\label{alg:bbb}
\SetAlgoLined
 init(MAX/ $P_0.ticket$...$P_n.ticket$/ $P_0.status$...$P_n.status$):= 
 	0/ 0...0/ noncrit...noncrit \;
 \While{true}{
 \ForEach{i in 0...n}{
  \uIf{select($P_i$)}{
   	$P_i.ticket$ = MAX + 1\;
   	$P_i.status$ = waiting\;
  	 }
   \uElseIf {$P_i.status$ = wait}{
   	\eIf{$P_i.ticket$ = min($P_0.ticket$... $P_n.ticket$)}{
		$P_i.status$ = crit \;
	}{
		$P_i.status$ = waiting \;
	}
      }}}
      
\end{algorithm}
}

\newcommand{\AlgSNARK}{
\begin{algorithm}[htb!]
\setcounter{AlgoLine}{0}
\caption{SNARK}
\label{alg:snark}
\SetAlgoLined
\SetKwFunction{Fun}{popRight}
\Fun{}{
\\ \While{true}{
	$rh$ = \textit{RightHat}\;
	$lh$ = \textit{LeftHat}\;
	\If{rh$ \rightarrow$R = rh}{
		return "empty"\;
	}
	\uIf{rh = lh}{
		\If{DCAS($\&$RightHat, $\&$LeftHat, rh, lh, Dummy, Dummy)}{
		return $rh$ $\rightarrow$$V$;}
	}
   	\uElse{
	rhL = rh$\rightarrow$L\;
   	\If{DCAS($\&$RightHat, $\&$rh$\rightarrow$L, rh, rhL, rhL, rh)}{
		\textit{result = rh}$\rightarrow$$V$\;
		$rh$$\rightarrow$$R$ = \textit{Dummy}\;
		return \textit{result}\;
	}
}}}

\SetKwFunction{Fun}{pushRight}
\Fun{}{
\\    	$nd$ = new Node()\;
	\If{nd = null}{
		return "full"\;
	}
	$nd$$\rightarrow$$R$ = \textit{Dummy}\;
	$nd$$\rightarrow$$V$ = $v$\;
	
 	\While{true}{	
	$rh$ = \textit{RightHat}s\;
	$rhR$ = $rh$$\rightarrow$$R$\;
	\uIf{rhR = rh}{
		$nd\rightarrow L$ = \textit{Dummy}\;
		$lh$ = \textit{LeftHat};\\
		\If{DCAS($\&$RightHat, $\&$LeftHat, rh, lh, nd, Dummy)}{
		return success\;}
	}
   	\uElse{
	$nd\rightarrow L$ = $rh$\;
   	\If{DCAS($\&$RightHat, $\&$lh$\rightarrow$R, rh, rhR, nd, nd)}{
	        return success\;
	}
}}}
\end{algorithm}
}

	


\newcommand{\AlgMultiThreaded}{
\begin{algorithm}[ht]
  \caption{Typed Multi-threaded Program}
  \label{alg:multithreaded}
\SetAlgoLined
 Thread $\alpha$:\\
 \While{mask != 0 }{
 	\While{trigger0 = 0}{no-op;}
	\textit{result} = \textit{result} $\|$ \textit{mask} ; // bitwise 'or' \\
	\textit{trigger0 = 0}\;
	\textit{maintrigger} = \textit{matintrigger + 1} \;
	\If{ maintrigger = 1}{
		\textit{trigger1} = $1$\;
	}
}
Thread $\beta$:\\
 \While{mask != 0 }{
 	\While{trigger1 = 0}{no-op;}
	\textit{result} = \textit{result} $\&$ !\textit{mask} ; // bitwise 'and' \\
	\textit{trigger1 = 0}\;
	\textit{maintrigger = matintrigger + 1} \;
	\If{ maintrigger = 1}{
		\textit{trigger0 = 1}\;
	}
}
Thread $\gamma$:\\
 \While{mask != 0 }{
 	\textit{maintrigger = 0} \;
	\eIf{PIN $\&$ mask = 0} {\textit{trigger0 = 1}\;}
	{ \textit{trigger1 = 1}\;}
 	\While{maintrigger != 2}{no-op;}
	\textit{mask = mask/2}\;
}
\textit{trigger0 = 1}\;
\textit{trigger1 = 1}\;
\end{algorithm}
}

\newcommand{\AlgNonRepudiation}{
\begin{algorithm}[]
  \caption{Non-repudiation Protocol}
  \label{alg:non-rep}
\begin{multicols}{2}
$T_{correct}$:\\
\hspace{0.6cm}(1) \textit{skip} until \textit{P$\rightarrow$T: m} \;
\hspace{0.6cm}(2) \textit{skip} until \textit{P$\rightarrow$T: NRO} \;
\hspace{0.6cm}(3) \textit{T}$\rightarrow$\textit{Q: m} \;
\hspace{0.6cm}(4) \textit{skip} until Q$\rightarrow$\textit{T: NRR}\;
\hspace{0.6cm}(5) \textit{T}$\rightarrow$\textit{Q: NRO}\;
\hspace{0.6cm}(6) \textit{T}$\rightarrow$\textit{P: NRR}\;

$T_{incorrect}$:\\
\hspace{0.6cm}(1) \textit{skip} until \textit{P}$\rightarrow$\textit{T: m} \;
\hspace{0.6cm}(2) \textit{skip} until \textit{P$\rightarrow$T: NRO} \;
\hspace{0.6cm}(3) \textit{T$\rightarrow$Q:m} \;
\hspace{0.6cm}(4) \textit{T$\rightarrow$Q:NRO}\;
\hspace{0.6cm}(5) \textit{skip} until \textit{Q$\rightarrow$T: NRR}s\;
\hspace{0.6cm}(6) \textit{T$\rightarrow$P:NRR}\;
\end{multicols}
\vspace{0.2cm}
\end{algorithm}
}

%% file: evals.tex
\section{Implementation and Empirical Evaluation}
\label{sec:evals}

\input{results} 

We have implemented the technique described in Section~\ref{sec:qbf}
in a tool called \HyperQube.
In this section, we describe this implementation and the empirical
evaluation of the case studies described in Section~\ref{sec:cases}.
The tool \HyperQube works as follows.
Given a transition relation, we automatically unfold it up to a given
bound $k \geq 0$ by a procedure {\em genqbf} using a home-grown tool
written in \Ocaml.

Given the choice of the semantics (pessimistic, optimistic,
h-pessimistic or h-optimistic) the unfolded transition relation is
combined with the QBF encoding of the input HyperLTL formula to form a
complete QBF instance which is then be fed to the state-of-the-art QBF
solver \code{Quabs}~\cite{ht18}.
All experiments in this section are run on an iMac desktop with Intel
i7 CPU @3.4 GHz and 32 GB of RAM.
%

\subsection{Evaluation of Case 1: Symmetry in the Bakery Algorithm}

The off-the-self Bakery algorithm described does not satisfy the
symmetry property, because when two or more process are intending to
enter the critical section with the same tickets number, the algorithm
always gives priority to the process with the smaller process ID.
We encode the Bakery program as Boolean formulas that encode the
initial states and the transition relation.
Then, we encoded the negation of the symmetry formula:
\begin{center}
\centering 
\begin{tabular}{ |c|c| } 
\hline
\thead{$\neg$Symmetry} & \thead{ $\neg \varphi_\sym = \exists \pi_{A}.  \forall 
\pi_{B}.\  {\F}\Big( \neg \sym(\select_{\pi_{A}}, \select_{\pi_{B}}) \lor  (\pause_{\pi_{A}} 
\neq \pause_{\pi_{B}})  \; \lor $ \\ $\hfill(\pc(P_0)_{\pi_{A}} \neq \pc(P_1)_{\pi_{B}}) 
\lor (\pc(P_1)_{\pi_{A}} \neq \pc(P_0)_{\pi_{B}})\Big)$} \\
\hline
\end{tabular}
\end{center}

\HyperQube returns SAT using the $\pes$ semantics, which indicates
that there exists a trace that satisfy $\neg \varphi_\sym$.
The returned trace represents a witness trace of Bakery that violates
symmetry and thus falsifies the original formula $\varphi_\sym$.

An observable witness within finite bound is sufficient with the 
\pes\ semantics to infer that all future observations are consistently
indicating the given model does not satisfy original property.

\subsection{Evaluation of Case  2: Linearizability in SNARK Algorithm}

The SNARK algorithm is not linearizable, which means that there is a
witness trace that has no sequential equivalent trace.
The violation of linearizability can be expressed as xthe negation of
the original property, as follows:

\begin{center}
\centering 
\begin{tabular}{ |c|c| } 
\hline
\thead{$\neg$ Linearizability} & \thead{ $\neg \varphi_{\linearizability} = 
\exists \pi_{A}. \forall \pi_{B}. \ {\F}(\history_{\pi_A} 
\not\leftrightarrow \history_{\pi_B})$} \\
\hline
\end{tabular}
\end{center}

In this case, \HyperQube returns SAT using the \pes semantics,
indicating that a witness of linearizability violation has been found.
Again, with the use of $\pes$ semantics, a witness of linearizability
violation of length $k$ is enough to infer that the given system does
not satisfy the linearizability property.
The bug we identified by using \HyperQube is the same as the bug trace
reported in~\cite{ddgflmmss04} with an ad-hoc technique.

\subsection{Evaluation of Case  3: Non-interference in Typed Multi-threaded Programs}

To verify non-interference, we use \HyperQube to search for a
counterexample exists.
We encode the following formula:

\begin{center}
\centering 
\begin{tabular}{ |c|c| } 
\hline
\thead{$\neg\NI$} & \thead{ $\neg\varphi_{\NI} = \exists \pi_{A}. \forall \pi_{B}. 
\big(\PIN_{\pi_{A}} \neq \PIN_{\pi_{B}} \big) \rightarrow \Big((\terminate_{\pi_{A}} \land \terminate_{\pi_{B}})$ \\
$\hspace{2cm} \mathcal R\ \big( (\neg \terminate_{\pi_{A}} \lor \neg 
\terminate_{\pi_{B}}) \lor  (\Result_{\pi_{A}}\neq \Result_{\pi_{B}})\big)\Big)\  $} \\
\hline
\end{tabular}
\end{center}

In this case we use \hpes to further exploit the terminating nature of
the system and, \HyperQube returns SAT, indicating that there is a
trace in which we can detect the difference of high-variable by
observing low variable, that is, violating non-interference.

\subsection{Evaluation of Case  4: Fairness in Non-repudiation Protocols}

In order to handle fairness in non-repudiation protocols we study the
negated formula, which is in $\forall\exists$ form against the
$T_{\incorrect}$ implementation.

\begin{center}
\centering 
\begin{tabular}{ |c|c| } 
\hline
\thead{$\neg$ Fairness} & \thead[l]{ $\neg\varphi_{\fair} = \forall \pi_{A}. \exists \pi_{B}.$ 
$\neg \big(({\F} m_{\pi_{A}}) \land ({\F} \NRR_{\pi_{A}}) \land ({\F}\NRO_{\pi_{A}})\big) \lor$ \\
$\hspace{1cm}  \Big(({\always} \bigwedge_{act \in \Act_P} act_{\pi_{A}} \leftrightarrow act_{\pi_{B}} )\ \land \neg \big(({\F} \NRR_{\pi_{B}}) \leftrightarrow ({\F} \NRO_{\pi_{B}})\big)\Big) \lor$\\
$\hspace{1cm}  \Big(({\always} \bigwedge_{act \in \Act_Q} act_{\pi_{A}} \leftrightarrow act_{\pi_{B}} )\ \land \neg \big(({\F} \NRR_{\pi_{B}}) \leftrightarrow ({\F} \NRO_{\pi_{B}})\big)\Big)$}\\
\hline
\end{tabular}
\end{center}

We obtain a SAT result from \HyperQube, but since the formula passed
to the solver is $\forall\exists$ the solver does not return an
witness.
Alternatively, one could verify the protocol with respect to formula
$\exists \pi_{A}. ({\F} m_{\pi_{A}} \land {\F} \NRR_{\pi_{A}} \land {\F}
\NRO_{\pi_{A}})$.
This step was successful, meaning that an effective trace exists,
meaning that the original SAT result implies that the protocol
includes an unfair trace.

We then studied the implementation named $T_{\correct}$
in~\cite{jmm11}, where $T$ always guarantees the message exchange
event is fair between the two parties.
In this case, \HyperQube returns UNSAT, which indicates that all
traces in the correct system satisfies fairness in non-repudiation.
In this case study, both SAT and UNSAT results from $\HyperQube$ can
be meaningful because of the use of halting semantics (\hpes\ for
falsification of $T_{\incorrect}$ and \hopt\ for verification of
$T_{\correct}$).

\subsection{Evaluation of Case  5: Privacy-Preserving Path Planning for Robots}

The use of HyperQube for robotic path planning is slightly different
from the above-mentioned cases.
In this case, we focus on synthesizing a qualified strategy that
satisfies the properties described above.
Thus, we enforce the original formulas including {\em shortest path}
and {\em robustness} properties directly with the map model.

\begin{itemize}
\item \textbf{Shortest path.} By encoding the map grid together with
  $\varphi_{\mathit{sp}}$, \HyperQube returns SAT.
  The returned path as shown in fig.~\ref{fig:robotic} represents a path
  that can reach the goal from the initial state with the least steps
  compared to all other paths.
\item \textbf{Robustness Path.} Encoding the map with ,
  $\varphi_{\mathit{rb}}$, \HyperQube again returns SAT.
  This corresponds to a robust strategy, in the sense that all other
  robots starting from an arbitrary initial state will eventually
  reach to the goal state by following exactly the same strategy.
  The result can be visualized in~\ref{fig:robotic_rb}
\end{itemize}

\input{roboticgraph}

We investigate the scalability and performance of our technique for
this particular study, in comparison with the technique introduced
in~\cite{wnp19}.
In~\cite{wnp19}, the paths for robotic planning are synthesized by 
unfolding the transition relations and properties using python scripts, 
and solve satisfiability using Z3 SMT solver~\cite{mb12}. 
The results shown in Table~\ref{Tab:robot} suggest that the QBF-based
approach of \HyperQube outperforms the solution
in~\cite{wnp19}---which is based on more mature SMT technology---, on
several challenging robotic planning problems.
As QBF solvers improve we anticipate that \HyperQube will
automatically benefit from their improvements.

\TableRobotPlanning

\subsection{Evaluation of Case 6: Generate Mutants in Mutation Testing}

We also evaluated HyperQube to synthesize valid mutants for mutation
testing as in~\cite{ftw19}.
We again directly apply the original formula that describes a good
mutant together with the model.
In this case, \HyperQube returns SAT, indicating that we have
successfully found a good qualified mutant.
Our experiment shows that \HyperQube is able to output a mutant with
the given formula in a very short amount of time, which provides an
efficient solution for test suite generation of mutation testing.

\subsection{Summary of Cases Results Evaluations}

Table \ref{Tab:results} shows the running times of \HyperQube in the
different case studies.

\TableResults

In Table~\ref{Tab:sems}, we separately address how we use
Theorem~\ref{thm:qbf-mc} to infer from the output of $\HyperQube$ the
result of the corresponding model-checking problem.

\TabCasesSemantics

The results shown in Table~\ref{Tab:sems} (\#0.1 to \#4.2) illustrate
that \HyperQube is capable of solving a variety of model checking
problems for alternating HyperLTL properties.
These instances are very challenging for techniques that attempt
reduce to model-checking of LTL because due to the complexity of
eliminating the alternation of quantifiers.
Additionally, QBF solvers allow to more efficient explore the search
space than a brute-force SAT-based approach, where universal and
existential quantifiers are eliminated by combinatorial expansion to
conjunctions and disjunctions.

In cases \#5.1 to \#6.2, we also demonstrate the ability of \HyperQube
to solve challenging synthesis problems by leveraging the existential
quantifier in a \HyperLTL formula as the synthesized result that
satisfies the specification.


%% file: results.tex
\newcommand{\TableRobotPlanning}{
  \begin{table}[ht]
\centering 
\begin{tabular}{|c | c |  c  c  c | c  c  c |} 
\cline{3-8}
  \multicolumn{2}{c}{}		     &	\multicolumn{3}{|c}{~\cite{wnp19}} & \multicolumn{3}{ | c |}{ \HyperQube } \\ [0.5ex]
  \hline\hline  
&    \#unroll & gen [s] & Z3 [s] & Total[s] & genqbf [s] & QuAbS [s] & Total [s] \\  [0.5ex]
\hline \hline
\thead{Shortest path \\ $(map\ size:  10^{2})$ } & 20 & 8.31 & 0.33 & \textbf{8.64} & 1.30 & 0.57 & \textbf{1.87}  \\ \hline
\thead{Shortest path \\ $(map\ size:  20^{2})$ } & 40 & 124.66 & 6.41 & \textbf{131.06} & 4.53 & 12.16 & \textbf{16.69}  \\  \hline
\thead{Shortest path \\ $(map\ size:  40^{2})$ } & 80 & 1093.12 & 72.99 & \textbf{1166.11} & 36.04 & 35.75 & \textbf{71.79}  \\  \hline
\thead{Shortest path \\ $(map\ size:  60^{2})$ } & 120 & 4360.75 & 532.11 & \textbf{4892.86} & 105.82 & 120.84 & \textbf{226.66}  \\  \hline\hline
\thead{Initial state robustness \\ $(map\ size:  10^{2})$ } & 20 & 11.14 & 0.45 & \textbf{11.59} & 1.40 & 0.35 & \textbf{1.75}  \\ \hline
\thead{Initial state robustness \\ $(map\ size:  20^{2})$ } & 40 & 49.59 & 2.67 & \textbf{52.26} & 15.92 & 15.32 & \textbf{31.14}  \\  \hline
\thead{Initial state robustness \\ $(map\ size:  40^{2})$ } & 80 & 216.16 & 19.81 & \textbf{235.97} & 63.16 & 20.13 & \textbf{83.29}  \\  \hline
\hline
\end{tabular}
\caption{Case studies results of hyperproperties for robotic planning on larger maps using 
\HyperQube, in comparison with the experimental results in ~\cite{cfst19}} 
\label{Tab:robot}
\end{table}
}

\newcommand{\TableResults}{
\begin{table}[ht]
\centering 
\begin{tabular}{|c |c | c c c c c c c  |} 
\hline\hline 
\# & Model  & $\varphi$ & \#unroll & QBF & sems & genqbf [s] & QuAbS [s] & Total [s]  \\ [0.5ex] 
\hline\hline 
0.1 & \thead{Bakery.3proc} 	& \thead{ $\forall\forall\ (\sym1)$} 	& 7  & SAT & $\textit{pes}$   & 0.44 & 0.04 & \textbf{0.48}  \\ 
\hline
0.2 &\thead{Bakery.3proc} 	& \thead{ $\forall\forall\ (\sym2)$} 	& 12 & SAT &  $\textit{pes}$   & 1.31 & 0.15&  \textbf{1.46}  \\ 
\hline
0.3 &\thead{Bakery.3proc} 	& \thead{ $\forall\forall\ (\sym3)$} 	& 20 & UNSAT & $\textit{opt}$  & 2.86 & 4.87 & \textbf{7.73}  \\ 
\hline
1.1 &\thead{Bakery.3proc} 	& \thead{ $\varphi_{\sym1}$ } 		& 10 & SAT &  $\textit{pes}$   & 0.86 & 0.11& \textbf{0.97}  \\ 
\hline
1.2 &\thead{Bakery.3proc} 	& \thead{ $\varphi_{\sym2}$ } 		& 10 & SAT & $\textit{pes}$  & 0.76 & 0.17&  \textbf{0.93} \\ 
\hline
1.3 &\thead{Bakery.5proc} 	& \thead{ $\varphi_{\sym1}$} 		& 10 & SAT & $\textit{pes}$   & 23.57 & 1.08 & \textbf{24.65} \\ 
\hline
1.4 &\thead{Bakery.5proc} 	& \thead{ $\varphi_{\sym2}$} 		& 10 & SAT & $\textit{pes}$   & 29.92 & 1.43&  \textbf{31.35} \\ 
\hline
2.1 &\thead{SNARK-bug1} 	& $\varphi_{\linearizability}$ 		& 26 & SAT & $\textit{pes}$ & 88.42  &  383.60&  \textbf{472.02}   \\
\hline
2.2 &\thead{SNARK-bug2} 	& $\varphi_{\linearizability}$ 		& 40 & SAT &$\textit{pes}$  & 718.09  &  779.76&  \textbf{1497.85}  \\
\hline
3.1 &\thead{3-Thread \\ ({\em incorrect})} &$\varphi_{\NI}$ 		& 57 & SAT & $\textit{h-pes}$ & 19.56 & 46.66&  \textbf{66.22}  \\
\hline
3.2 &\thead{3-Thread \\ ({\em correct)}} &$\varphi_{\NI}$ 			& 57 & UNSAT & $\textit{h-opt}$  & 23.91 & 33.54 & \textbf{57.45}  \\
\hline
4.1 &\thead{NRP\\ ($T_{\incorrect}$)} & $\varphi_{\fair}$ 			& 15 & SAT & $\textit{h-pes}$   & 0.10 & 0.27&  \textbf{0.37} \\
\hline
4.2 &\thead{NRP\\ ($T_{\correct}$)}& $\varphi_{\fair}$ 			& 15 & UNSAT & $\textit{h-opt}$   & 0.08 & 0.12 & \textbf{0.20} \\
\hline
5.1 &\thead{Shortest Path}& $\varphi_{\mathit{sp}}$ & 20  & SAT  & $\textit{h-pes}$ & 1.30 & 0.57 & \textbf{1.87} \\
\hline
5.2 &\thead{Initial State\\ Robustness}& $\varphi_{\mathit{rb}}$ & 20 & SAT   & $\textit{h-pes}$   & 1.40 & 0.35 & \textbf{1.75} \\
\hline
6.1 &\thead{Mutant\\ Synthesis}& $\varphi_{\mut}$ & 20 & SAT   & $\textit{h-pes}$   & 1.40 & 0.35 & \textbf{1.75} \\
\hline
\end{tabular}
\caption{Performance of \HyperQube in the case studies. 
Column {\em case\#} identifies the artifact, and the rest of the
columns represent the models, properties, number of unrolling in BMC,
semantic used for infinite inference, and the running time for
generating the query and for solving it.} 
\label{Tab:results}
\end{table}
}

\newcommand{\TabCasesSemantics}{
\begin{table}[ht]
\centering 
\begin{tabular}{|c |c |c |c |c | l |} 
\hline\hline 
Semantics  & Case\# & Property & QBF & Infinite Inference & Conclusion    \\ [0.5ex] 
\hline\hline
\multirow{2}{*}{\pes} 	& \thead{0.1\\ 0.2\\ 1.1\\ 1.2\\ 1.3\\ 1.4} &  $\varphi_{\sym}$ 	
					& SAT & \multirow{4}{*}{\thead[c]{$\Kr \Pkmodels \neg \varphi$$\so \Kr \models \neg \varphi$}} 
					& $\Kr$  $\not\models$  $\varphi_{\sym}$ \\\cline{2-4}\cline{6-6}
					& \thead{2.1\\ 2.2} &  $\varphi_{\linearizability}$ 	
					& SAT &  
					& $\Kr$ $\not\models$   $\varphi_{\linearizability}$   \\[1ex]\hline\hline
\opt 				&\thead{0.3} &  $\varphi_{\sym}$ 	
					& UNSAT & \thead[c]{$\Kr \Okmodels \neg \varphi$$\so \Kr \not\models \neg \varphi$} 
					& $\Kr$ $\models$  $\varphi_{\sym}$ \\[1ex]\hline\hline
\multirow{2}{*}{$h$-\pes}	& \thead{3.1} 	& $\varphi_{\NI}$ 	
					& SAT & \multirow{2}{*}{\thead[c]{$\Kr \HPkmodels \neg \varphi$$\so \Kr \models \neg \varphi$}} 
					& $\Kr$ $\not\models$  $\varphi_{\NI}$ \\\cline{2-4}\cline{6-6}
					& \thead{4.1}					& $\varphi_{\fair}$   	
					& SAT &  
					& $\Kr$ $\not\models$ $\varphi_{\fair}$\\[1ex]\hline\hline
\multirow{2}{*}{$h$-\opt}	& \thead{3.2} 	& $\varphi_{\NI}$ 	
					& UNSAT & \multirow{2.5}{*}{\thead[c]{$\Kr \not\HOkmodels \neg \varphi$$\so \Kr \not\models \neg \varphi$}} 
					& $\Kr$ $\models$ $\varphi_{\NI}$ \\\cline{2-4}\cline{6-6}
					& \thead{4.2}						& $\varphi_{\fair}$   	
					& UNSAT &  
					& $\Kr$ $\models$ $\varphi_{\fair}$\\[1ex]\hline\hline
\multirow{4}{*}{\thead[c]{Synthesis \\  (\pes)}} 	& \thead{5.1} & $\varphi_{\mathit{sp}}$ 		
									& SAT & \multirow{4.5}{*}{\thead[c]{$\Kr \Pkmodels \varphi$$\so \Kr \models \varphi$}} 
									& shortest path exists\\[1ex]\cline{2-4}\cline{6-6}
									& \thead{5.2} & $\varphi_{\mathit{rb}}$ 			
									& SAT &  
									& robust path exists\\\cline{2-4}\cline{6-6}
									& \thead{6.1} & $\varphi_{\mut}$ 			
									& SAT &  
									& mutant synthesized\\
\hline\hline
\end{tabular}
\caption{Mappings of cases studies and model checking problem conclusions, 
with different semantics used for infinite inference from Theorem~\ref{thm:qbf-mc}.} 
\label{Tab:sems}
\end{table}
}

%% file: roboticgraph.tex
\begin{figure}[!htb]
\minipage{0.46\textwidth}
\centering
\begin{adjustbox}{width=3.8cm}
\begin{tikzpicture}  
    [
        box/.style={rectangle,draw=gray,thick, minimum size=1cm}, 
    ]

\foreach \x in {0,1,...,9}{
    \foreach \y in {0,1,...,9}
        \node[box] at (\x,\y){};
}

\node[box,fill=red  ] at (0,0){}; 
\node[box,fill=green] at (7,5){};  

\draw[arw, ->, line width=0.5mm, color = blue] (0,0) -- (0,1);
\draw[arw, ->, line width=0.5mm, color = blue] (0,1) -- (0,2);
\draw[arw, ->, line width=0.5mm, color = blue] (0,2) -- (0,3);
\draw[arw, ->, line width=0.5mm, color = blue] (0,3) -- (0,4);
\draw[arw, ->, line width=0.5mm, color = blue] (0,4) -- (0,5);
\draw[arw, ->, line width=0.5mm, color = blue] (0,5) -- (0,6);
\draw[arw, ->, line width=0.5mm, color = blue] (0,6) -- (0,7);
\draw[arw, ->, line width=0.5mm, color = blue] (0,7) -- (1,7);
\draw[arw, ->, line width=0.5mm, color = blue] (1,7) -- (2,7);
\draw[arw, ->, line width=0.5mm, color = blue] (2,7) -- (3,7);
\draw[arw, ->, line width=0.5mm, color = blue] (3,7) -- (4,7);
\draw[arw, ->, line width=0.5mm, color = blue] (4,7) -- (5,7);
\draw[arw, ->, line width=0.5mm, color = blue] (5,7) -- (5,6);
\draw[arw, ->, line width=0.5mm, color = blue] (5,6) -- (5,5);
\draw[arw, ->, line width=0.5mm, color = blue] (5,5) -- (6,5);
\draw[arw, ->, line width=0.5mm, color = blue] (6,5) -- (7,5);

\node[box,fill=black] at (5,0){};
\node[box,fill=black] at (7,0){};
\node[box,fill=black] at (3,1){};
\node[box,fill=black] at (9,1){};
\node[box,fill=black] at (1,3){};
\node[box,fill=black] at (3,3){};
\node[box,fill=black] at (6,3){};
\node[box,fill=black] at (3,4){};
\node[box,fill=black] at (5,4){};
\node[box,fill=black] at (6,4){};
\node[box,fill=black] at (7,4){};
\node[box,fill=black] at (1,5){};
\node[box,fill=black] at (3,5){};
\node[box,fill=black] at (4,5){};
\node[box,fill=black] at (8,5){};
\node[box,fill=black] at (1,6){};
\node[box,fill=black] at (2,6){};
\node[box,fill=black] at (3,6){};
\node[box,fill=black] at (4,6){};
\node[box,fill=black] at (6,6){};
\node[box,fill=black] at (7,6){};
\node[box,fill=black] at (8,6){};
\node[box,fill=black] at (1,8){};
\node[box,fill=black] at (2,8){};
\node[box,fill=black] at (3,8){};
\node[box,fill=black] at (1,9){};
\node[box,fill=black] at (2,9){};
\node[box,fill=black] at (4,9){};

\end{tikzpicture}%
\end{adjustbox}%
  
  \caption{Shortest Path}
  \label{fig:robotic}
\endminipage\hfill
\minipage{0.46\textwidth}
  \centering
\begin{adjustbox}{width=3.8cm}
\begin{tikzpicture} 
    [
        box/.style={rectangle,draw=gray,thick, minimum size=1cm}, 
    ]

\foreach \x in {0,1,...,9}{
    \foreach \y in {0,1,...,9}
        \node[box] at (\x,\y){};
}

\node[box,fill=red  ] at (0,0){};
\node[box,fill=red  ] at (1,0){};
\node[box,fill=red  ] at (2,0){};
 
\node[box,fill=green] at (6,9){};  
\node[box,fill=green] at (7,9){};  
\node[box,fill=green] at (8,9){};  

\draw[arw, ->, line width=0.5mm, color = blue] (2,0) -- (3,0);
\draw[arw, ->, line width=0.5mm, color = blue] (3,0) -- (3,1);
\draw[arw, ->, line width=0.5mm, color = blue] (3,1) -- (4,1);
\draw[arw, ->, line width=0.5mm, color = blue] (4,1) -- (4,2);
\draw[arw, ->, line width=0.5mm, color = blue] (4,2) -- (5,2);
\draw[arw, ->, line width=0.5mm, color = blue] (5,2) -- (5,3);
\draw[arw, ->, line width=0.5mm, color = blue] (5,3) -- (6,3);
\draw[arw, ->, line width=0.5mm, color = blue] (6,3) -- (6,4);
\draw[arw, ->, line width=0.5mm, color = blue] (6,4) -- (6,5);
\draw[arw, ->, line width=0.5mm, color = blue] (6,5) -- (6,6);
\draw[arw, ->, line width=0.5mm, color = blue] (6,6) -- (6,7);
\draw[arw, ->, line width=0.5mm, color = blue] (6,7) -- (7,7);
\draw[arw, ->, line width=0.5mm, color = blue] (7,7) -- (7,8);
\draw[arw, ->, line width=0.5mm, color = blue] (7,8) -- (7,9);
\draw[arw, ->, line width=0.5mm, color = blue] (7,9) -- (8,9);

\node[box,fill=black] at (5,0){};
\node[box,fill=black] at (6,0){};
\node[box,fill=black] at (7,0){};
\node[box,fill=black] at (8,0){};
\node[box,fill=black] at (9,0){};
\node[box,fill=black] at (6,1){};
\node[box,fill=black] at (7,1){};
\node[box,fill=black] at (8,1){};
\node[box,fill=black] at (9,1){};
\node[box,fill=black] at (7,2){};
\node[box,fill=black] at (8,2){};
\node[box,fill=black] at (9,2){};
\node[box,fill=black] at (8,3){};
\node[box,fill=black] at (9,3){};
\node[box,fill=black] at (9,4){};

\node[box,fill=black] at (0,5){};
\node[box,fill=black] at (0,6){};
\node[box,fill=black] at (1,6){};
\node[box,fill=black] at (0,7){};
\node[box,fill=black] at (1,7){};
\node[box,fill=black] at (2,7){};
\node[box,fill=black] at (0,8){};
\node[box,fill=black] at (1,8){};
\node[box,fill=black] at (2,8){};
\node[box,fill=black] at (3,8){};
\node[box,fill=black] at (0,9){};
\node[box,fill=black] at (1,9){};
\node[box,fill=black] at (2,9){};
\node[box,fill=black] at (3,9){};
\node[box,fill=black] at (4,9){};

\end{tikzpicture}
\end{adjustbox}
  \caption{Robust Strategy}
 \label{fig:robotic_rb}
\endminipage
\end{figure}

%% file: related.tex
\section{Related Work}
\label{sec:related}


There has been a lot of recent progress in automatically
verifying~\cite{frs15,fmsz17,fht18,cfst19} and
monitoring~\cite{ab16,fhst19,bsb17,bss18,fhst18,sssb19,hst19} HyperLTL
specifications.
HyperLTL is also supported by a growing set of tools, including the
model checker MCHyper~\cite{frs15,cfst19}, the satisfiability checkers
EAHyper~\cite{fhs17} and MGHyper~\cite{fhh18}, and the runtime
monitoring tool RVHyper~\cite{fhst18}.

The complexity of the \emph{model checking} for HyperLTL for
tree-shaped, acyclic, and general graphs was rigorously investigated
in~\cite{bf18}.
The first algorithms for model checking HyperLTL and HyperCTL$^*$
using alternating automata were introduced in~\cite{frs15}.
These techniques, however, were not able to deal in practice with
alternating HyperLTL formulas in a fully automated fashion.
We also note that previous approaches that reduce model checking
HyperLTL---typically of formulas without quantifier alternations---to
model checking LTL can use BMC in the LTL model checking phase.
However, this is a completely different approach than the one
presented here, as these approaches simply instruct the model checker
to use a BMC \emph{after} the problem has beenfully reduced to an LTL
model checking problem while we avoid this translation.

These algorithms were then extended to deal with hyperliveness and
alternating formulas in~\cite{cfst19} by finding a winning strategy in
$\forall\exists$ games.
In this paper, we take an alternative approach by reducing the model
checking problem to QBF solving, which is arguably more effective for
finding bugs (in case a finite witness exists).

The {\em satisfiability} problem for HyperLTL is shown to be
undecidable in general but decidable for the $\exists^*\forall^*$
fragment and for any fragment that includes a $\forall\exists$
quantifier alternation~\cite{fh16}.
The hierarchy of hyperlogics beyond HyperLTL were studied
in~\cite{cfhh19}.
The synthesis problem for HyperLTL has been studied in problem
in~\cite{bf19} in the form of {\em program repair}, in~\cite{bf20} in
the form of {\em controller synthesis}, and in~\cite{fhlst20} for the
general case.


%% file: concl.tex
\section{Conclusion and Future Work}
\label{sec:concl}

In this paper, we introduced the first bounded model checking (BMC)
technique for verification of hyperproperties expressed in
HyperLTL.
To this end, we proposed four different semantics that ensure the
soundness of inferring the outcome of the model-checking problem.
To handle trace quantification in HyperLTL, we reduced the BMC problem
to checking satisfiability of quantified Boolean formulas (QBF).
This is analogous to the reduction of BMC for LTL to the simple
propositional satisfiability problem.
We have introduced different classes of semantics, beyond the
pessimistic semantics common in LTL model checking, namely optimistic
semantics that allow to infer full verification by observing only a
finite prefix and halting variations of these semantics that
additionally exploit the termination of the execution, when available.

Through a rich set of case studies, we demonstrated the effectiveness
and efficiency of our approach in verification of information-flow
properties, linearizability in concurrent data structures, path
planning in robotics, and fairness in non-repudiation protocols.

As for future work, our first step is to solve the loop condition
problem.
This is necessary to establish completeness conditions for BMC and can
help cover even more examples efficiently.
The application of QBF-based techniques in the framework of
abstraction/refinement is another unexplored area.
Success of BMC for hyperproperties inherently depends on effectiveness
of QBF solvers.
Even though QBF solving is not as mature as SAT/SMT solving
techniques, recent breakthroughs on QBF have enabled the construction
of \HyperQube, and more progress in QBF solving will improve its
efficiency.